\documentclass[intlimits,twoside,a4paper]{article}
\usepackage[cp1251]{inputenc}

\usepackage[eqsecnum]{cmpj3}

\usepackage{bm}
\usepackage{soul,xcolor}
\setstcolor{red}

\newcommand{\mycomment}[1]{}

\issue{2025}{28}{4}{43704}
\doinumber{10.5488/CMP.28.43704}

\title[Link of the \textit{Zitterbewegung} with the  spin conductivity and the spin-textures of multiband systems]
{Link of the \textit{Zitterbewegung} with the  spin conductivity and the spin-textures of multiband systems}

\author[F. Mireles, E. Ortiz]
{F. Mireles\orcid{0000-0003-0506-0793}\refaddr{label1}\thanks{Corresponding author: \email{fmireles@ens.cnyn.unam.mx}}, 
{E. Ortiz\orcid{0009-0008-8348-0519}
        \refaddr{label1,label2}}}
     
\addresses{
\addr{label1} Departamento de F\'isica, Centro de Nanociencias y Nanotecnolog\'ia, Universidad Nacional Aut\'onoma de M\'exico (UNAM),
Apartado Postal 14, 22800 Ensenada, Baja California, Mexico
\addr{label2}Department of Electrical Engineering, University of Notre Dame, Notre Dame, IN 46556, USA
}


\date{Received August 23, 2025, in final form October 28, 2025}
\begin{document}
\maketitle

\begin{abstract}
The \textit{Zitterbewegung} phenomenon in multiband electronic systems is known to be subtly related to the charge conductivity, Berry curvature and the Chern number.
Here we show that some spin-dependent properties as the optical spin conductivity and intrinsic spin Hall conductivity are also entangled with the  \textit{Zitterbewegung} amplitudes. We also show that in multiband Dirac-type Hamiltonians, a direct link between the \textit{Zitterbewegung} and the spin textures and spin transition amplitudes can be established. The later allow us to discern the presence or not of the   \textit{Zitterbewegung} oscillations by simply analyzing the spin or pseudospin textures. We provide examples of the applicability of our approach for Hamiltonian models that show the suppression of specific  \textit{Zitterbewegung} oscillations. 
\printkeywords 
\end{abstract}

\section{Introduction}

Schr\"odinger \cite{schrodinger1930kraftefreie} predicted in 1930 that a quantum relativistic free particle following Dirac's Hamiltonian will develop an oscillatory behavior of high frequency, producing a trembling motion (\textit{Zitterbewegung}) as it travels, causing rapid oscillations in the particle's
location and speed. The trembling behavior results as consequence the coupling of particle and antiparticle states of the Dirac Hamiltonian. 
Today it is understood that the \textit{Zitterbewegung} phenomenon
is not strictly a relativistic effect, as it arises also in any quantum physical system that couples the linear momentum of the quasiparticles with its physical spin or its pseudospin degrees of freedom \cite{GeneralTheory-PRB2010,CertiDavid2010}. Earlier studies of \textit{Zitterbewegung} in solids were performed in 1970 by Lurie and Cremer \cite{Lurie&Cremer} in superconductor materials, and two decades later by Cannata et al.~\cite{Cannata} in bulk semiconductor materials. The topic was then essentially forgotten for more than a decade when,  with the advent of the semiconductor spintronics field \cite{SpintronicsReview,Hirohata}, it showed a revival in 2005 with the work of Schliemann \cite{schliemann2005zitterbewegung}, and independently by Jiang et al. \cite{jiang2005semiclassical} and  Zawadaski \cite{Zawadski-PRB2005}, putting forward the plausibility for the observation of the \textit{Zitterbewegung} in two-dimensional electron and hole systems with sizable spin-orbit coupling. 

Nevertheless, the very first experimental measurement of the \textit{Zitterbewegung} oscillations was realized until 2010 in the realm of trapped ions by Gerritsma et al. \cite{gerritsma2010quantum} utilizing counter-incident lasers to construct a synthetic Dirac system based on an ultra cold $^{40}$Ca$^+$ gas.  Soon after, the \textit{Zitterbewegung} effect was also confirmed in spin-orbit-coupled Bose-Einstein condensates of ultracold $^{87}$Rb atoms \cite{leblanc2013direct,SOCBEC}. 
Recently, the \textit{Zitterbewegung} oscillations were also observed in planar photonic microcavities 
and in analogues of honeycomb graphene coupled to microcavity resonators \cite{OpticalSimulationZitter}. Even more recently (2024),  the experimental observation  of trembling quantum motion (\textit{Zitterbewegung}) of exciton polaritons in a perovskite microcavity was reported to occur at room temperature \cite{perovskite}.

The \textit{Zitterbewegung} effect has also been explored theoretically in a number of two-dimensional materials: in graphene \cite{Maksimova,RusinZawadski,RusinZawadski2008,Carrillo2018}, Kekul\'e distorted graphene \cite{A.SantaCruz}, phosphorene \cite{wpd-phosphorene}, 
silicene \cite{wpd-silicene,wpd-silicene2,Hassan}, and
borophene \cite{borophene2}. It was also studied in dice lattices ($\alpha-T_3$ models) \cite{Biswas}, topological insulators \cite{wpd-ti,wpd-ti1,wpd-ti2}, 
%and even more recently, a study of the moir\'{e} excitons dynamics in van der Waals heterostructures of MoS$_2$/WSe$_2$ have shown to exhibit a trembling motion \cite{PeetersgroupPRL2021}. 
Recently, the study of the dynamics of moir\'{e} excitons in van der Waals heterostructures of
MoS$_2$/WSe$_2$ has shown that they exhibit a trembling motion \cite{PeetersgroupPRL2021}.

A general theory of the \textit{Zitterbewegung} dynamics that has been put forward by D\'avid and Cserti 
\cite{GeneralTheory-PRB2010} is applicable for any $\bm k$-dependent multiband electronic system. The central idea is that the \textit{Zitterbewegung} could arise in any given $N$-band Hamiltonian
if there exists a lock (coupling) of the orbital motion (momentum) to real or virtual (pseudo) spin of the quasiparticle.  In this work the authors derive alternative expressions for the time dependence of the position operator in terms of nondiagonal elements of the \textit{Zitterbewegung} amplitudes, denoted by $\hat{\boldsymbol{Z}}_{ab}$  ($a\neq b$),  $a,\,b$ being the band index. Very interestingly, in a follow up work by the same  authors \cite{CertiDavid2010}, there was found a direct link between the charge conductivity and  the \textit{Zitterbewegung} amplitudes,  via the so-called Berry connection matrix that enters the canonical expression for the Berry phase \cite{Berry}. Furthermore, they show that the Berry curvature and Chern number are related to the diagonal elements $\hat{\boldsymbol{Z}}_{aa}$. These seminal contributions show that the \textit{Zitterbewegung} is indeed interconnected to other measured physical quantities. 

A crucial aspect of this general framework for the \textit{Zitterbewegung} is the explicit appearance of a multifrequency beating in the time dependence of the position and velocity operators associated with the quasiparticles. Such frequencies correspond to all possible energy differences of the eigenvalues of the multiband Hamiltonian, $\omega_{ab}=(E_a -E_b)/\hbar$. For instance, the two-band Rashba spin-orbit system~\cite{Rashba,Bychkov, Santana,Bercioux}, the single-layer graphene, and the (heavy and light) hole Luttinger model \cite{Luttinger}, described all by $2\times2$ Hamiltonians,  have trivially only one beating frequency. On the other hand, bilayer graphene which is described by a $4\times4$ spinless Hamiltonian, has two electron and two hole bands, which accounts in total with six energy differences, and naturally, with only four different oscillation frequencies. 

However, in a recent study of the \textit{Zitterbewegung} in the graphene with a Y-shaped Kekul\'e bond texture~\cite{A.SantaCruz} we show that certain optical transitions (beating frequencies) are found to be suppressed.
Since the low-energy electronic excitations in the graphene with a 
Kekul\'e-Y distortion are described by a four-band Hamiltonian \cite{Gamayun}  --- due to its valley-momentum and pseudospin-momentum locking nature --- we would then expect oscillations of four different frequencies in the \textit{Zitterbewegung} dynamics, as it occurs in the bilayer graphene. By contrast, only two frequencies are shown to be present. The non-contributing aspect related to the \textit{Zitterbewegung} oscillations was attributed to  the vanishing of the Berry connection matrix elements for the involved 
states and, alternatively, to the symmetry arguments. 
Intriguingly, these states show a unique pattern of its {valley} and {pseudospin textures}. 

In this work, we show that the \textit{Zitterbewegung} amplitudes in multiband electronic systems are also linked to spin-transport and magnetic properties, as the spin conductivity and intrinsic spin Hall conductivity. 
We further show that the spin (pseudospin) textures, as well as the off-diagonal spin (pseudospin) matrix elements of general two-dimensional (2D) spin-orbit coupled (SOC) systems and Dirac-type Hamiltonians are directly connected with the presence or absence of the \textit{Zitterbewegung} oscillations in such systems.

\section{{\textit Zitterwebegung}: general theory}

The D\'avid-Cserti's general theory of the \textit{Zitterbewegung} motion \cite{GeneralTheory-PRB2010} applies for an arbitrary multiband Hamiltonian ${\hat H}({\bm k})$, where ${\bm k}$ denotes the wave number of its Bloch states. The central idea is first to decompose the Hamiltonian in the form $\hat{H} = \sum_a{E_a(\boldsymbol{k})\hat{Q}_a(\boldsymbol{k})}$, where $E_a(\boldsymbol{k})$ are its \textit{a-th}  eigenvalue and $\hat{Q}_a(\boldsymbol{k})=| \psi_a({\bm k})\rangle\langle \psi_a({\bm k})|$ are the projection operators associated to the eigenkets $|\psi_a({\bm k})\rangle$. Since the $n\times n$ Hamiltonian is Hermitian,  then $\hat{Q}_a(\boldsymbol{k})\hat{Q}_b(\boldsymbol{k}) = \delta_{ab}|\psi_a(\boldsymbol{k})\rangle\langle \psi_a(\boldsymbol{k})|$, and  $\sum_a{\hat{Q}_a(\boldsymbol{k})} = I_n$ are satisfied, where $I_n$ is the $n\times n$ identity matrix, while $n$ means the band-index. Then, the position operator $ \hat{\boldsymbol{x}}(t)$ written in the Heisenberg picture can be rewritten as \cite{CertiDavid2010}, 

\begin{equation}
\hat{\boldsymbol{x}}(t) = \hat{\boldsymbol{x}}_o(0) +  \hat{\boldsymbol{\xi}}(t), \quad \hat{\boldsymbol{x}}_o(0) = \hat{\boldsymbol{x}}(0)+ \sum_{a} \hat{\boldsymbol{Z}}^{aa} ,
\end{equation}

%\begin{equation} \hat{\boldsymbol{x}}(t) = \hat{\boldsymbol{x}}_o(0) +  \hat{\boldsymbol{\xi}}(t) \end{equation}

 %\noindent where, the time independent part is

%\begin{equation}    \hat{\boldsymbol{x}}_o(0) = \hat{\boldsymbol{x}}(0)+ \sum_{a} \hat{\boldsymbol{Z}}^{aa}  \end{equation}

\noindent in which $\hat{\boldsymbol{x}}(0)$ is the position operator at $t=0$ (Schr\"odinger picture) and the time dependent term is given by

\begin{equation}\label{Oscillatory}
\hspace{1cm}  \hat{\boldsymbol{\xi}}(t) = t\sum_a{\boldsymbol{V}_a \hat{Q}_a } + \sum_a\sum_{b\neq a}{\re^{\ri\omega_{ab}t}\hat{\boldsymbol{Z}}^{ab} },
\end{equation}
 where $\boldsymbol{V}_a(\boldsymbol{k})=\frac{1}{\hbar}\frac{\partial E_a}{\partial {\bm k}}$ is the quasiparticle velocity, $\omega_{ab}=(E_a -E_b)/\hbar$ are the interband {\textit{Zitterwebegung}} frequency beatings, and its amplitudes are defined by 
\begin{equation}\label{Zab}
  \hat{\boldsymbol{Z}}^{ab} = \ri\hat{Q}_a \frac{\partial \hat{Q}_b}{\partial {\bm k}} .
\end{equation}

From the oscillatory part of  equation~(\ref{Oscillatory}) it is noted that a multi-frequency {\textit{Zitterwebegung}} could indeed arise in multiband systems, and that the presence of the oscillations are governed solely by the interband amplitudes $\hat{\boldsymbol{Z}}_{ab}$. Therefore, a necessary condition for the existence of {\textit{Zitterwebegung}} beatings is that amplitudes $\hat{\boldsymbol{Z}}_{ab}\ne 0$; while evidently, for the cases in which $\hat{\boldsymbol{Z}}_{ab}=0$, entails directly a total absence of {\textit{Zitterwebegung}} oscillations. Physically, the  $\hat{\boldsymbol{Z}}_{ab}$ is also linked to the Berry connection matrix, 
\begin{equation}\label{eq:Berryconnection}
    \boldsymbol{A}_{ab}(\boldsymbol{k}) = \ri\langle \psi_a(\boldsymbol{k})|\frac{\partial}{\partial \boldsymbol{k}}|\psi_b(\boldsymbol{k})\rangle ,
\end{equation} 

\noindent since $\langle \psi_a(\boldsymbol{k})|\hat{\boldsymbol{Z}}^{ab}|\psi_b(\boldsymbol{k})\rangle\equiv \langle a |\hat{\boldsymbol{Z}}^{ab}| b\rangle=\boldsymbol{A}_{ab}(\boldsymbol{k})$, which enters in the definition of Berry curvature~\cite{Berry}. 

\section{{\textit Zitterwebegung} amplitudes link to the spin conductivity }

In order to seek for a relation of the  {\textit Zitterwebegung} amplitudes with the frequency dependent spin conductivity, as well as with the intrinsic spin Hall conductivity,  we start by considering the general correlation function  $\Pi_{AB}$ for any two operators $A$ and $B$.  As dictated by the Kubo linear response formula, and following D\'avid and Cserti \cite{GeneralTheory-PRB2010},  once the correlation function is written in terms of the projectors operators $\hat Q_a$,  it reads,%(Appendix \ref{A1})  
\begin{equation}\label{Pi}
   \Pi_{AB}(\ri\nu_m)=\frac{1}{V}\sum_{{\bm k}}\sum_{a,b} K_{ba}(\ri\nu_m)\rm{Tr}\left[A \hat Q_a B \hat Q_b\right],
\end{equation}

\noindent where $V$ is the volume of the system, $\nu_m=2\piup m/\beta$ are the bosonic Matsubara's frequencies, $m$ being an integer, and 
\begin{equation} \label{SHCond} 
K_{ab}(\ri\nu_m) = \frac{n_F(E_b - \mu) - n_F(E_a - \mu)}{E_b - E_a + \ri\nu_m},
\end{equation}

\noindent in which  $\mu$ is the chemical potential, $n_F(E) =(1+\re^{\beta E})^{-1}$ is the Fermi-Dirac distribution, and $\beta=1/k_{\text{B}} T$ with $T$ being the temperature. Therefore, for the
spin-current--charge-current correlation function [$Q_{ij}^l(\ri\nu_m) \equiv \Pi_{{\cal J}_{i}J_j}(\ri\nu_m)$] we have,
\begin{equation}
    Q_{ij}^l(\ri\nu_m)=\frac{1}{V}\sum_{{\bm k}}\sum_{a,b} K_{ba}(\ri\nu_m) {\rm Tr}\left[{\cal J}_i^l \hat Q_a J_j \hat Q_b\right], 
\end{equation}

\noindent which is associated with a driven spin-current ${\cal J}_i^l$ polarized along the $l$-direction as a response to an applied electric field $E_j$ and a charge-current $J_i$. The spin-current operator is defined by  
${\cal J}_i^l= \frac{1}{4}\{\sigma_l,\frac{\partial H}{\partial k_i}\}$, and the charge current operator is given by  $J_j= ev_j=\frac{e}{\hbar}\frac{\partial H}{\partial k_j} $. 
The frequency dependent spin conductivity \cite{Wong} is then calculated after analytic continuation $\ri\nu_m \rightarrow \hbar\omega + \ri\delta $\footnote{In \cite{Bernevig} the energies are measured in units $\hbar^2$, consequently its equation~(6) for the spin conductivity should be divided by $\hbar^2$ to get the right units of $e/\hbar$ for the spin conductivity.}  as \cite{Bernevig}%,Footnote}
\begin{equation}\label{SpinC}
    \sigma_{ij}^l(\omega) = -\frac{1}{\hbar^2}\lim_{\delta \rightarrow 0^+}  \frac{Q_{ij}^l(\omega +\ri\delta)}{{i\omega}} .
\end{equation}

Using equation~(\ref{Zab}) for the definition of \textit{Zitterbewegung} amplitude, together with the identity, $
\hat Q_a \frac{\partial H}{\partial k_i} \hat Q_b = \delta_{ab} \frac{\partial E_a}{\partial k_i} \hat Q_a + (E_b - E_a) \hat Z_i^{ab}$ for $a\neq b$,  and the definitions of charge and spin current operators, it is straightforward to show that  the spin-current--charge-current correlation function can be rewritten as,

\begin{equation}\label{Qij}
   Q_{ij}^l(\ri\nu_m) = -\frac{e}{2 \ri\hbar}\frac{1}{V}\sum_{\mathbf{k}}\sum_{a,b, a \neq b} K_{ba} (\ri\nu_m)   (E_a - E_b) \text{Tr}  \left [ \frac{\hbar}{2} \left \{ \sigma_l , v_i\right\}   \hat Z_j^{ab}  \right ].
 \end{equation}

 \noindent Therefore, given that $ Q_{ij}^l(i\nu_m)$ depends explicitly on $\hat Z_j^{ab}$,
 from equation~(\ref{SpinC}) we can see that the optical (frequency-dependent) spin conductivity is  directly connected to the \textit{Zitterbewegung} phenomena, similarly as it occurs with the charge-conductivity \cite{CertiDavid2010}. Interestingly,  as a consequence, note that no spin-current is expected associated with the optical transitions between bands $E_a$ and $E_b$ whenever $\hat Z_i^{ab}=0.$ 

Now, using equations~(\ref{SpinC}) and (\ref{Qij}) the intrinsic spin Hall conductivity for a two dimensional system in the static limit  $\omega \rightarrow 0$ is given by (Appendix \ref{Eq3.6Derivation}), 

\begin{equation}\label{Sigma0}
\sigma_{ij}^l(0) = \frac{e}{\hbar A} \sum_{\mathbf{k}} \sum_{a, a \neq b} \frac{n_F(E_a) - n_F(E_b)}{(E_a - E_b)} \text{Tr}  \left [ \frac{\hbar}{2} \left \{ \sigma_l , v_i\right\}   \hat Z_j^{ab}  \right ],
\end{equation}

\noindent where $A$ is the area of the sample. 
%Hence, also the dc intrinsic spin Hall conductivity depend in general on the {\textit Zitterwebegung} amplitudes, and evidently vanishes if $\hat Z_i^{ab}=0.$ 
Hence, the direct current (dc) intrinsic spin Hall conductivity also depends in general on the Zitterwebegung amplitudes and evidently vanishes if $\hat Z_i^{ab}=0$ .
Note that equation~(\ref{Sigma0}) reduces to the known Kubo formula for the static spin Hall conductivity \cite{Guo,Feng}.

\section{Connection of \textit{Zitterbewegung}  with the spin, pseudospin and valley isospin textures}
\label{Fifth}

In order to analyze further the connection of the \textit{Zitterbewegung} with spin and/or pseudospin related properties,  such as the spin, pseudospin and valley isospin textures, we begin by considering a generic multiband Hamiltonian of the form,  
\begin{equation}\label{eq:con}
    H(\bm k) =h(\boldsymbol{\bm k})+\sum_{\eta=1}^{N_s} \alpha_\eta \left(\bm k \cdot \boldsymbol{\mathcal S}_{\eta}\right),
    %,\hspace{0.5cm} \eta =\{s,\sigma,\tau}\}
\end{equation}

%\noindent that might contain  a spin, pseudospin, and valley isopin -momentum  locking terms.  
\noindent where $h(\bm{k})$ stands for a (possible) second order in $\bm{k}$ dependence of the carriers, which is assumed not to be coupled to other degrees of freedom. The second expression to the right  represents the locking of the linear momentum with the spin, lattice pseudospin, valley isospin, or any other degrees of freedom~(${\boldsymbol{\mathcal{S}}}_{\eta}$).  The coupling constant is $\alpha_\eta$, $N_s$ is the number of terms that contain {``spin''} degrees of freedom that might be present in the model Hamiltonian. For instance, in a two-dimensional electron gas (2DEG) with Rashba spin-orbit coupling (SOC), the two-band model Hamiltonian $h(\bm k)= \hbar^2 k^2/2m$, with $\bm k =(k_x,k_y)$ and has only one spin degree of freedom~($N_s =1$), namely the physical spin,  $\alpha_{1}=\alpha_{\text{R}}$ is the Rashba coupling parameter, and ${\boldsymbol{\mathcal{S}}}_{1}=(-s_y,s_x)$, where $s_{x,y}$ being are the Pauli spin matrices (times  $\hbar/2$) describing the components of the physical spin $\bm s$. On the other hand, for the four band low energy Hamiltonian describing the low energy excitations in graphene monolayer near the Dirac points ($K,K'$), we have instead $h(\bm k)=0$, $N_s=2$, ${\boldsymbol{\mathcal{S}}}_{1}=(\mathcal{S}^x_1,0)$, ${\boldsymbol{\mathcal{S}}}_{2}=(0,\mathcal{S}^y_1)$, with   $\mathcal{S}^x_1=\tau_z \otimes \sigma_x$, $\mathcal{S}^y_1=\tau_o \otimes \sigma_y$, and coupling parameter $\alpha_1 =\alpha_2 = \hbar v_F$, where $v_F$ is the Fermi velocity.
In this way, the general expression in (\ref{eq:con})
comprises broad types of effective Hamiltonians that models a wide type of materials, ranging from 2DEG systems with Rashba and/or Dresselhaus spin-orbit coupling, to two dimensional graphene, Weyl semimetals and topological insulators among others. In table \ref{2DHams} we show some 2D Hamiltonians with its explicit form for the $\boldsymbol{{S}}_\eta$. 
%To easy the nomenclature, for now on, ann at least explicitly especified, 
\begin{table}[htbp]
	\caption{Explicit form for $\boldsymbol{\mathcal{S}}_{\eta}$ for several 2D Hamiltonians. Here, the physical spin is ${\bm s}=(s_x,s_y,s_z)$,  ${\bm \sigma}=(\sigma_x,\sigma_y)$ is the sublattice pseudospin, ${\bm \tau}=(\tau_x,\tau_y)$ is the valley isospin, all in terms of the Pauli matrices, $\sigma_0,\tau_0$ are $2\times2$ unit matrices, and $S_i$ with $i=\{x,y,z\}$ are the components of the $3\times3$ spin-1 matrices when $\alpha=1$. }
	\vspace{0.2cm}
\centering
\begin{tabular}{|l||c|c|c|}
\hline
\hline
\textbf{2D Hamiltonian} & coupling(s) $\alpha_{\eta}$ & $\boldsymbol{\mathcal{S}}_{\eta}$ &  explicit value for $\boldsymbol{\mathcal{S}}_{\eta}$\\
\hline
\hline
Rashba SOC            & $\alpha_{R}$     & $\boldsymbol{\mathcal{S}}_{R}$   & $\left(-s_y, s_x\right)$ \\
\hline
Dresselhaus SOC       & $\alpha_{D}$     & $\boldsymbol{\mathcal{S}}_{D}$   & $\left(s_x, -s_y\right)$ \\
\hline
Rashba + Dresselhaus ( $\alpha_{\text{R}}=\alpha_{\text{D}}$ ) & $ \beta$ & $\boldsymbol{\mathcal{S}}_{R}+ \boldsymbol{\mathcal{S}}_{D}$ & $\left(s_x-s_y, s_x-s_y\right)$ \\
\hline
Graphene $K$            & $\hbar v_F$      & $\boldsymbol{\mathcal{S}}_{K}$   & $\left(\sigma_x, \sigma_y\right)$ \\
%\hline
%Chern insulator       & $\delta$         & $\boldsymbol{\mathcal{S}}_{\text{CI}}$ & $\left(\sigma_x, \sigma_y\right)$ \\
\hline
Graphene $K'$         & $\hbar v_F$      & $\boldsymbol{\mathcal{S}}_{K'}$  & $\left(-\sigma_x, \sigma_y\right)$ \\
\hline
Full Graphene $K,K'$         & $\hbar v_F$      & $\boldsymbol{\mathcal{S}}_{1}$, $\boldsymbol{\mathcal{S}}_{2}$  & $\left(\tau_z \otimes \sigma_x,0 \right), (0,\tau_0 \otimes \sigma_y)$ \\
\hline
Graphene Kek Y        & $v_{\sigma}, v_{\tau}$ & $\boldsymbol{\mathcal{S}}_{\sigma}, \boldsymbol{\mathcal{S}}_{\tau}$ & $\tau_0 \otimes \boldsymbol{\sigma}, \; \boldsymbol{\tau} \otimes \sigma_0$ \\
\hline
$\alpha-T_3$ dice lattice       & $v_F$ & $\boldsymbol{{S}}$ & $(S_x,S_y,S_z)$ \\
%\hline
%Graphene + Rashba     & $\hbar v_F$      & $\boldsymbol{\mathcal{S}}_{\sigma}$ & $\tau_0 \otimes \boldsymbol{\sigma}$ \\
%\hline 
% Graphene Kek Y + Rashba SOC & $\hbar v_{\sigma}, \hbar v_{\tau}$ & $\boldsymbol{\mathcal{S}}_{\sigma}, \boldsymbol{\mathcal{S}}_{\tau}$ & $\tau_0 \otimes \boldsymbol{\sigma} \otimes s_0, \; \boldsymbol{\tau} \otimes \sigma_0 \otimes s_0$ \\
\hline
\end{tabular}
\label{2DHams}
\end{table}

 A general connection of the \textit{Zitterbewegung} amplitudes with the spin, pseudospin, valley isospin textures, and with the interband spin transitions can be found for such multiband Hamiltonians~(\ref{eq:con}). Here, we show (Appendix \ref{Eq5.2Derivation}) that such a connection must follow (for $a \neq b$) 
\begin{equation}\label{eq:proof}
\begin{aligned}
       \langle a |\hat{\boldsymbol{Z}}^{ab}| b\rangle  &=  \frac{1}{2\hbar \omega_{b a}}\sum_{\eta=1}^{N_s} \alpha_{\eta} \biggl(\langle{\boldsymbol{\mathcal{S}}}_{\eta}\rangle_{a a}+ \langle{\boldsymbol{\mathcal{S}}}_{\eta}\rangle_{b b}+2\ri\langle{\boldsymbol{\mathcal{S}}}_{\eta}\rangle_{a b}\biggr) \\
   &+\frac{1}{\hbar \omega_{b a}}\nabla_{\boldsymbol{k}} \biggl(h(\boldsymbol{k})-\frac{1}{2}\hspace{-1em}\sum_{m \in\{a, b\}}E_m({\bm k})\biggr),
   \end{aligned}
\end{equation}

 %$\langle i\nabla_{\boldsymbol{k}}\rangle_{a b}$
\noindent where the diagonal matrix elements $\langle{\boldsymbol{\mathcal{S}}}_{\eta}\rangle_{a a}\equiv\langle \psi_a(\boldsymbol{k})|\boldsymbol{\mathcal S}_{\eta}|\psi_a(\boldsymbol{k})\rangle$ define the ``spin'' orientation or the average  ``spin'' in a given band,
%and momentum $\bm k$
while the off-diagonal matrix elements
$\langle{\boldsymbol{\mathcal{S}}}_{\eta}\rangle_{a b}$, describe the interband spin ``{\textit texture}'' or spin transition amplitudes. Note that for any Hamiltonian having  $h(\boldsymbol{k}) = 0$, as the Dirac-type and low energy Weyl Hamiltonians, a vanishing contribution of the last term in~(\ref{eq:proof}) is always yielded owing to the electron-hole symmetry of such systems, hence   

%then and only then, $2h(\boldsymbol{k})-\sum_{m \in\{a b\}}E_m = 0$,  hence,

\begin{equation}\label{eq:proofmir}
    \langle a |\hat{\boldsymbol{Z}}^{ab}| b\rangle  =  \frac{1}{4E_b} \sum_\eta\alpha_\eta\biggl(\langle {\boldsymbol{\mathcal{S}}}_{\eta}\rangle_{a a}+ \langle {\boldsymbol{\mathcal{S}}}_{\eta}\rangle_{b b}+2\ri\langle {\boldsymbol{\mathcal{S}}}_{\eta}\rangle_{a b}\biggr) =\boldsymbol{A}_{ab}(\boldsymbol{k}) . 
\end{equation}

This expression yields then an alternative formula for the Berry connection matrix $\boldsymbol{A}_{ab}(\boldsymbol{k})$ in such systems. Interestingly, it involvs just the inter- and intra-band (pseudo)spin-textures. Hence, whenever a zero  output is produced from the term in parenthesis in (\ref{eq:proofmir}),
%involving the sum of the matrix elements of the $\boldsymbol{\mathcal{S}}_{\eta}$, 
it entails that the Berry connection matrix $\boldsymbol{A}_{ab}(\boldsymbol{k})=0$,  which in turn directly implies the absence of a \textit{Zitterbewegung} oscillation with frequency $\omega_{ab}$. In other words, from equation (\ref{eq:proofmir}) it  follows that whenever
\begin{equation}\label{ZBcondition}
   \langle{\boldsymbol{\mathcal{S}}}_{\eta}\rangle_{a a}+ \langle {\boldsymbol{\mathcal{S}}}_{\eta}\rangle_{b b}+2\ri\langle {\boldsymbol{\mathcal{S}}}_{\eta}\rangle_{a b}  = 0
\end{equation}

%\begin{equation}  \sum_\eta\alpha_\eta\left(\langle {\boldsymbol{\mathcal{S}}}^{\eta}\rangle_{a a}+ \langle {\boldsymbol{\mathcal{S}}}^{\eta}\rangle_{b b}+2i\langle {\boldsymbol{\mathcal{S}}}^{\eta}\rangle_{a b}\right)   = 0 \end{equation}

\noindent is satisfied for each degree of freedom $\eta$,  a condition for \textit{Zitterbewegung} forbidden transitions between the given bands $E_a$ and $E_b$ is established. Note that the restriction of considering \textit{Zitterbewegung} transitions only between non-degenerate energy states is not a significant limitation. As first elucidated by Cserti and D\'avid, this theoretical framework is inherently designed for non-degenerate states, as \textit{Zitterbewegung} naturally vanishes when transitions occur between degenerate states \cite{GeneralTheory-PRB2010}. We remark that  condition~(\ref{ZBcondition}) holds in general for systems following the generic Hamiltonian (\ref{eq:con}) in which the whole term $h(\bm k)-\frac{1}{2}\sum_{m}E_{m}$ or its $\bm k$-gradient vanishes. Further simplifications can be obtained for the expression~(\ref{ZBcondition}) if symmetry arguments are considered. 
For instance, the Rashba Hamiltonian  that satisfies both,  the lack of space-inversion  and time-reversal symmetry, always produces opposite (pseudo)spin-textures for different bands, $\langle{\boldsymbol{\mathcal{S}}}_{\eta}\rangle_{a a}=- \langle {\boldsymbol{\mathcal{S}}}_{\eta}\rangle_{bb}$ (figure~\ref{fig:SRD2}). However, it can be shown that $\langle{\boldsymbol{\mathcal{S}}}_{\eta}\rangle_{a b}\ne 0$ for the Rashba Hamiltonian, and hence the condition  (\ref{ZBcondition}) is not  satisfied; which in turn is consistent with  the (nonzero) frequency \textit{Zitterbewegung} oscillations expected in this system \cite{schliemann2005zitterbewegung}. 

We now proceed to examine the absence of \textit{Zitterbewegung} oscillations of particular frequencies  in specific Hamiltonian systems in connection with the derived general condition (\ref{ZBcondition}). Three illustrative examples are considered, (i) the joint Rashba and Dresselhaus spin-orbit Hamiltonian in 2DEGs, (ii) graphene with Kekul\'e distortion, and (iii) the $\alpha-T_3$ lattice model Hamiltonian.

\subsection{The Rashba-Dresselhaus Hamiltonian}

Let us consider a two dimensional electron system under the simultaneous presence of Rashba and (linear) Dresselhaus spin-orbit interactions \cite{Rashba,Bychkov,Santana,Bercioux,Dresselhaus}. Its Hamiltonian written in the notation of (\ref{eq:con}) with $N_s =1$, $\alpha_1=\alpha_{\text{R}}$, $\alpha_2=\alpha_{\text{D}}$ reads, 
 
\begin{equation}
   H_{\text{RD}}(\bm k)= \frac{\hbar^2k^2}{2m^*}+\alpha_{\text{R}} (\bm k\cdot {\boldsymbol{\mathcal S}_{\text{R}}})+\alpha_{\text{D}} (\bm k\cdot {\boldsymbol{\mathcal S}_{\text{D}}}),
\end{equation}

 %\begin{equation}\label{eq:rdh}     H_{\text{RD}}(\bm k)= \frac{\hbar^2k^2}{2m^*}+\alpha\left( k_y {s}_x-k_x {s}_y\right)+ \beta \left( k_y {s}_y-k_x \hat{s}_x\right) \end{equation}

\noindent with ${\bm k}=(k_x,k_y)$ is the in-plane electron wave-vector, $\alpha_{\text{R}}$ is the Rashba coupling parameter (which is gate-tunable), and $\alpha_{\text{D}}$  is the Dresselhaus coupling parameter. Here, ${\boldsymbol{\mathcal S}_{\text{R}}}=(-s_y,s_x,0)$  and ${\boldsymbol{\mathcal S}_{\text{D}}}=(-s_x,s_y,0)$. Our  particular interest is the case of equal strength  of the Rashba and Dresselhaus coupling parameters ($\alpha_{\text{R}} = \alpha_{\text{D}}\equiv \beta$) in which Schliemann et al. \cite{schliemann2005zitterbewegung} first identified the absence of \textit{Zitterbewegung} oscillations. In this case we have
\begin{equation}
   H_{\text{RD}}(\bm k)= \frac{\hbar^2k^2}{2m^*}+\beta \bm k\cdot ({\boldsymbol{\mathcal S}_{\text{R}}}+{\boldsymbol{\mathcal S}_{\text{D}}})
\end{equation}

\noindent leading to the spin-splitting of the electron bands in momentum space and eigenvectors,   

\begin{equation}
    \mathcal{E}_{\pm}({\bm{k}})=\frac{\hbar^2k^2}{2m^*}\pm\sqrt{2}\beta|k_x+k_y| ,  \,\quad
    |{\bm k},\pm\rangle=\frac{{\rm e}^{\ri{{\bm k}}\cdot\mathbf{r}}}{\sqrt{2}}\left(\begin{array}{cc}
      1   \\
      \mp {\rm{e}}^{\ri{\piup}/{4}}  
    \end{array}\right)
    \label{eigenvalor_equal_rashba_dresselhaus}.
\end{equation}

%\begin{equation}
%    E_{\pm}(\bm k) = \frac{\hbar^2 k^2}{2m^*}\pm \sqrt{\left(\alpha_{\text{R}} k_x- \alpha_{\text{D}} k_y\right)^2+\left(\alpha_{\text{R}} k_y- \alpha_{\text{D}} k_x\right)^2} \end{equation}

\begin{figure}[!t]
    \centering
%     \hspace{-0.7cm}
 \includegraphics[width=4.0cm,height=6.0cm]{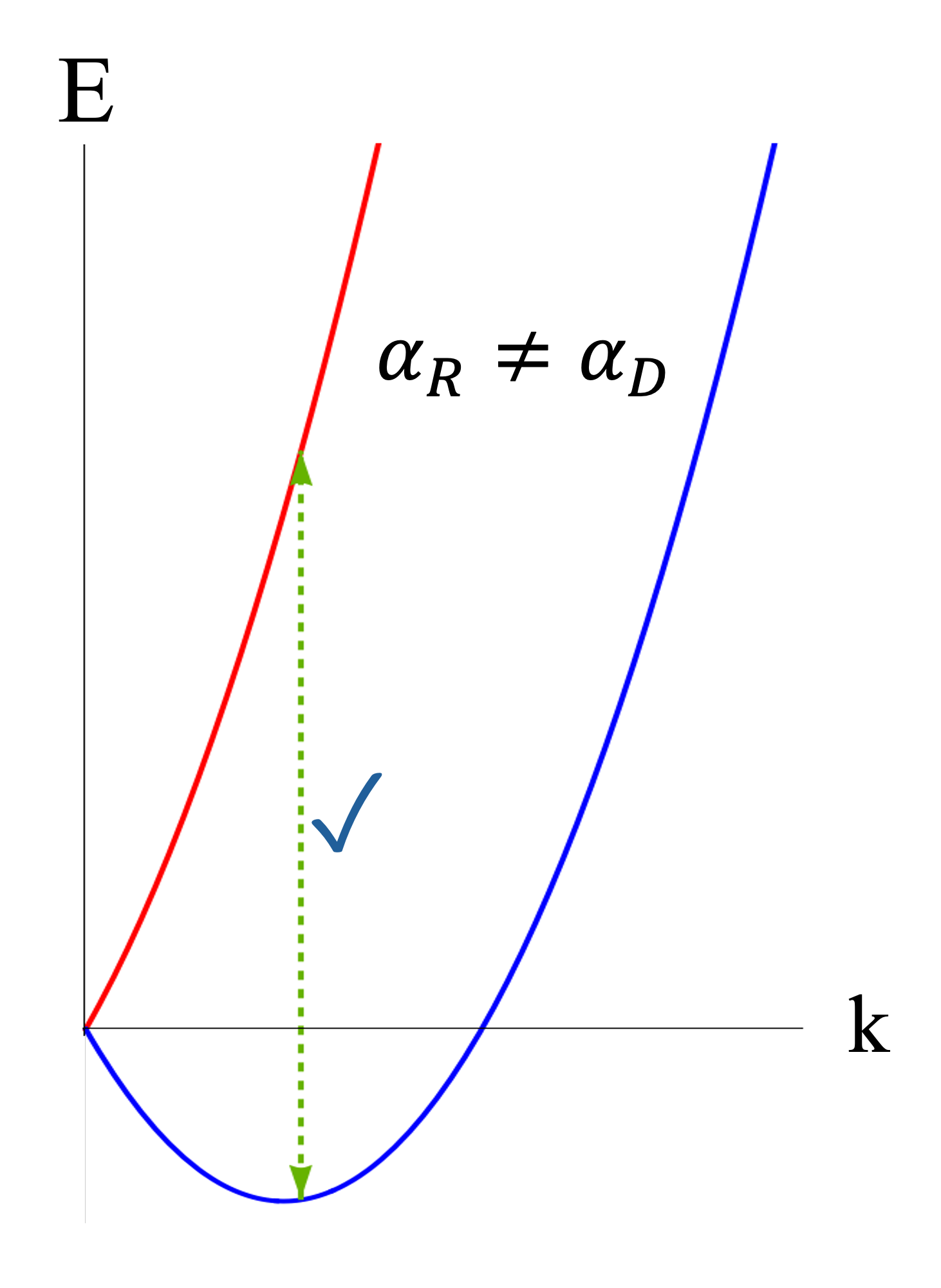}
  \hspace{2.4cm}
 \includegraphics[width=4.0cm,height=6.0cm]{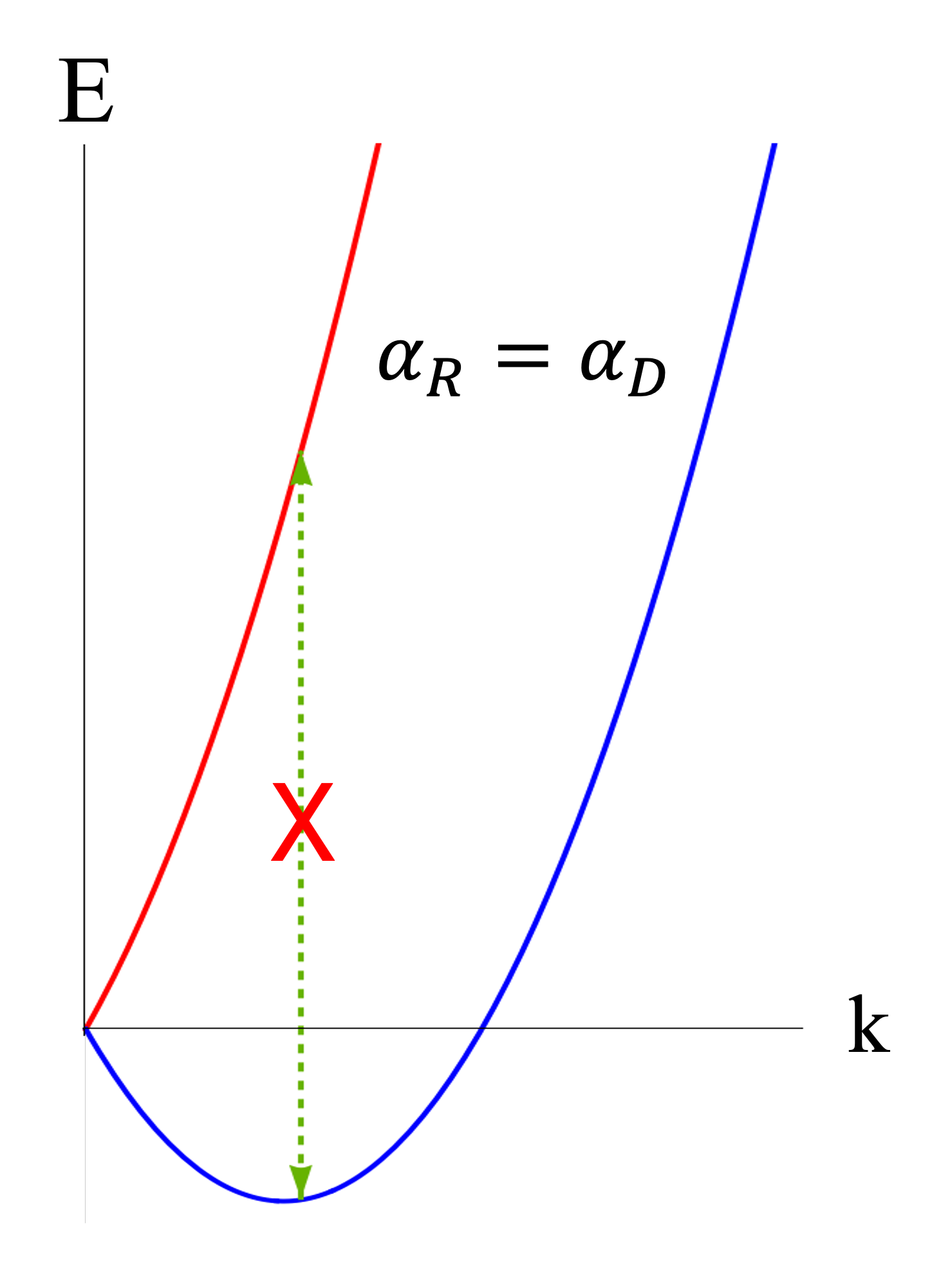}
   \hspace{0.7cm}\vspace{0.75cm}
   \includegraphics[width=6.5cm,height=5.0cm]{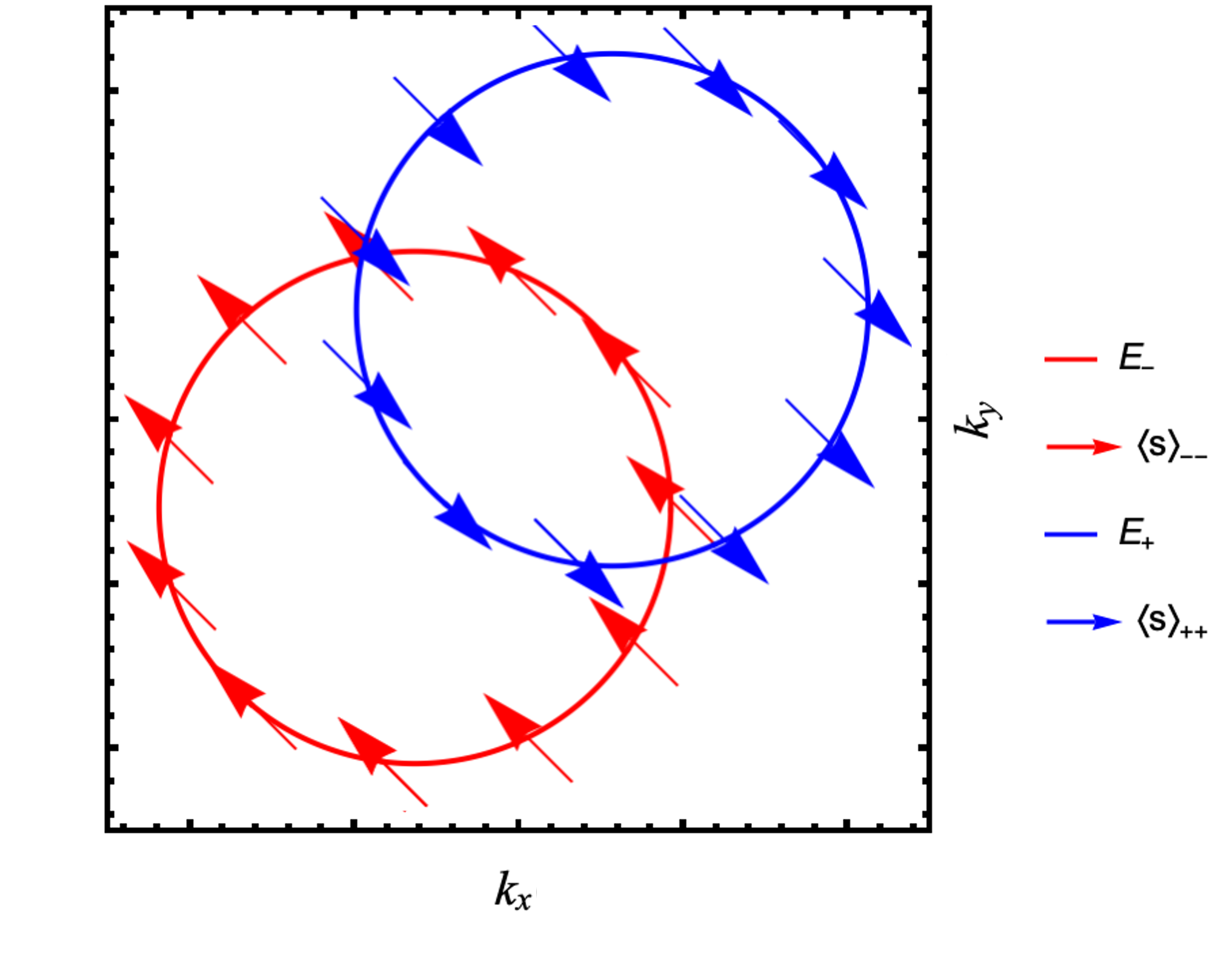}
 \includegraphics[width=7.0cm,height=5.0cm]{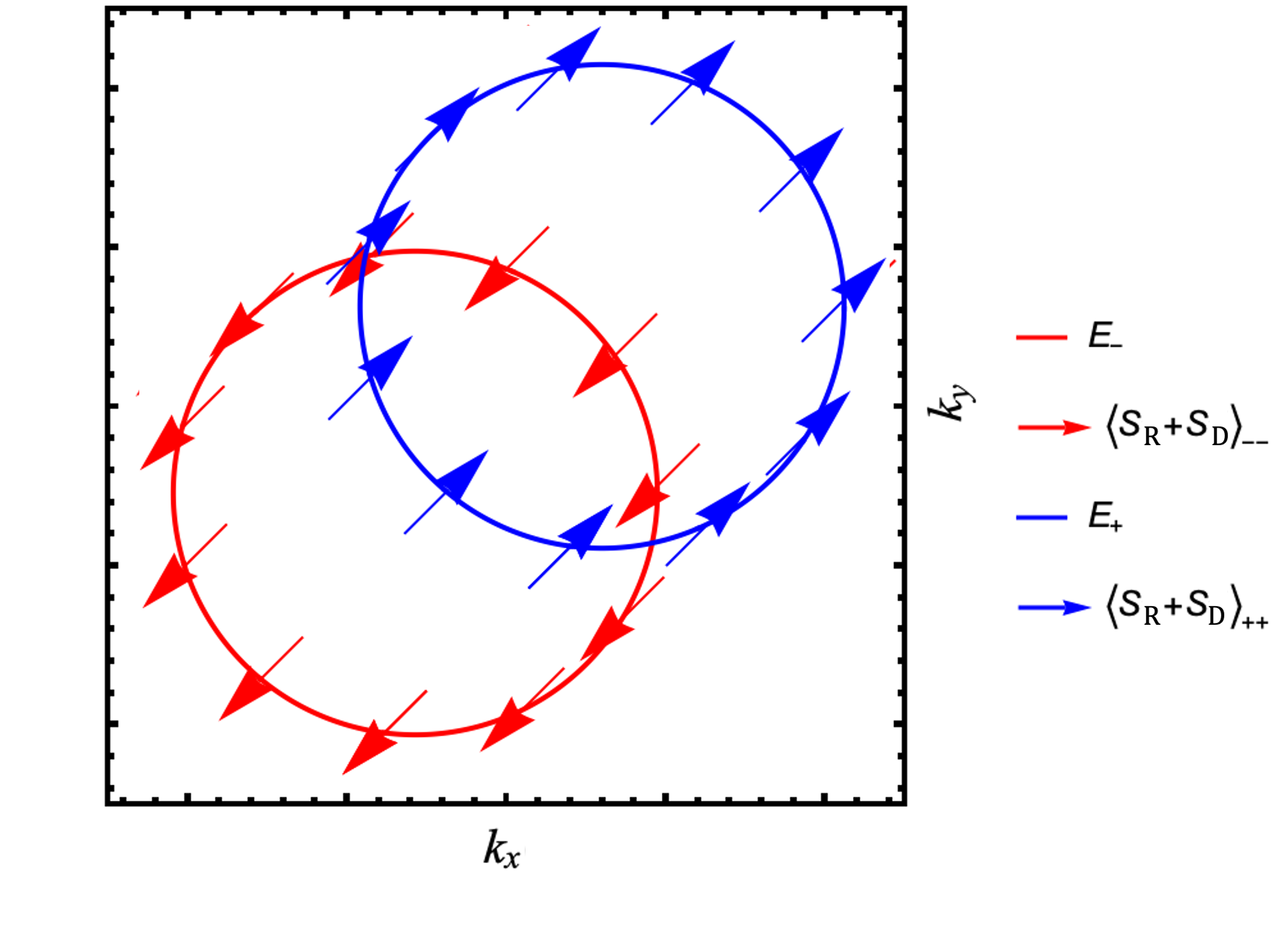}
  \\
 \vspace{-0.8cm}
    \caption{(Colour online) (Above) Energy dispersions of the Rashba and Dresselhaus Hamiltonian. (Top left-hand) with different spin-orbit coupling strengths ($\alpha_{\text{R}} \neq\alpha_{\text{D}}$), 
    (top right-hand) with equal  spin-orbit coupling strengths ($\alpha_{\text{R}} =\alpha_{\text{D}}$). The symbols $\surd$ in blue and $\bf \times$ in red indicate an allowed and an inhibited  \textit{Zitterbewegung} frequency  (spin-transition), respectively. (Bottom) Schematic diagram illustrating the contour plots of the ${\cal E}_-$ and  ${\cal E}_+$, both for $\alpha_{\text{R}} =\alpha_{\text{D}}$ at a fix energy $E$; (left) depicting the spin-texture $\langle \bm s\rangle $, and (right-hand) with depiction of the pseudospin texture $\langle \boldsymbol{{\mathcal{S}}}_{\text{R}}+ \boldsymbol{{\mathcal{S}}}_{\text{D}}\rangle$ for each band.}
    \label{fig:SRD2} 
\end{figure}

\noindent For this particular Hamiltonian,  the standard spin-texture notion defined as $\langle \bm s \rangle_{\pm}= \langle{\bm k},\pm |{\bm s}|{\bm k},\pm\rangle$, which yields $\pm\tfrac{1}{\sqrt{2}}(1,-1,0)$, differs from the pseudospin texture defined $\langle \mathcal{\bm S}_{\eta} \rangle_{\pm}= \langle{\bm k},\pm |\mathcal{\bm S}_{\eta}|{\bm k},\pm\rangle$,  which for this case results in $ \langle{\bm k},\pm |\boldsymbol{\mathcal S}_{\text{R}}+{\boldsymbol{\mathcal S}_{\text{D}}}|{\bm k},\pm\rangle=\pm {\sqrt{2}}(1,1,0)$.   
The electron energy dispersion and the spin and pseudospin textures are shown  in  figure~\ref{fig:SRD2}. Now, note that 

\begin{equation}
   \nabla_{\boldsymbol{k}}\left(\frac{\hbar^2k^2}{2m^*}-\frac{1}{2}( \mathcal{E}_{+} + \mathcal{E}_-)\right) = 0, 
\end{equation}

\noindent which together with the results $\langle + |{\boldsymbol{\mathcal S}_{\text{R}}}+{\boldsymbol{\mathcal S}_{\text{D}}})|+\rangle=-\langle - |{\boldsymbol{\mathcal S}_{\text{R}}}+{\boldsymbol{\mathcal S}_{\text{D}}})|-\rangle$ and  $\langle \pm |\boldsymbol{{\mathcal{S}}}_{\text{R}}|\mp\rangle = \langle \pm |\boldsymbol{{\mathcal{S}}}_{\text{D}}|\mp\rangle = 0$, yelds the condition (\ref{ZBcondition}) is thus fully satisfied. This ensures the absence of \textit{Zitterbewegung} oscillation with frequency $\omega_{+-}$ whenever the situation of a joint Rashba-Dresselhaus SOC is present with equal values of its coupling strengths. That is, without the need of calculating the equations of motion, for the position or velocity operators, we can directly infer using (\ref{ZBcondition}) that for the case $\alpha_{\text{R}} = \alpha_{\text{D}}$ no \textit{Zitterbewegung} oscillation will occur. The vertical green arrows in figure~\ref{fig:SRD2} represents the prohibited transition frequency. 
 
\subsection{The Kekul\'e distorted graphene Hamiltonian}
\begin{figure}[!t]
    \centering
 \vspace{-0.1cm}
     \includegraphics[width=6.5cm,height=5.5cm]{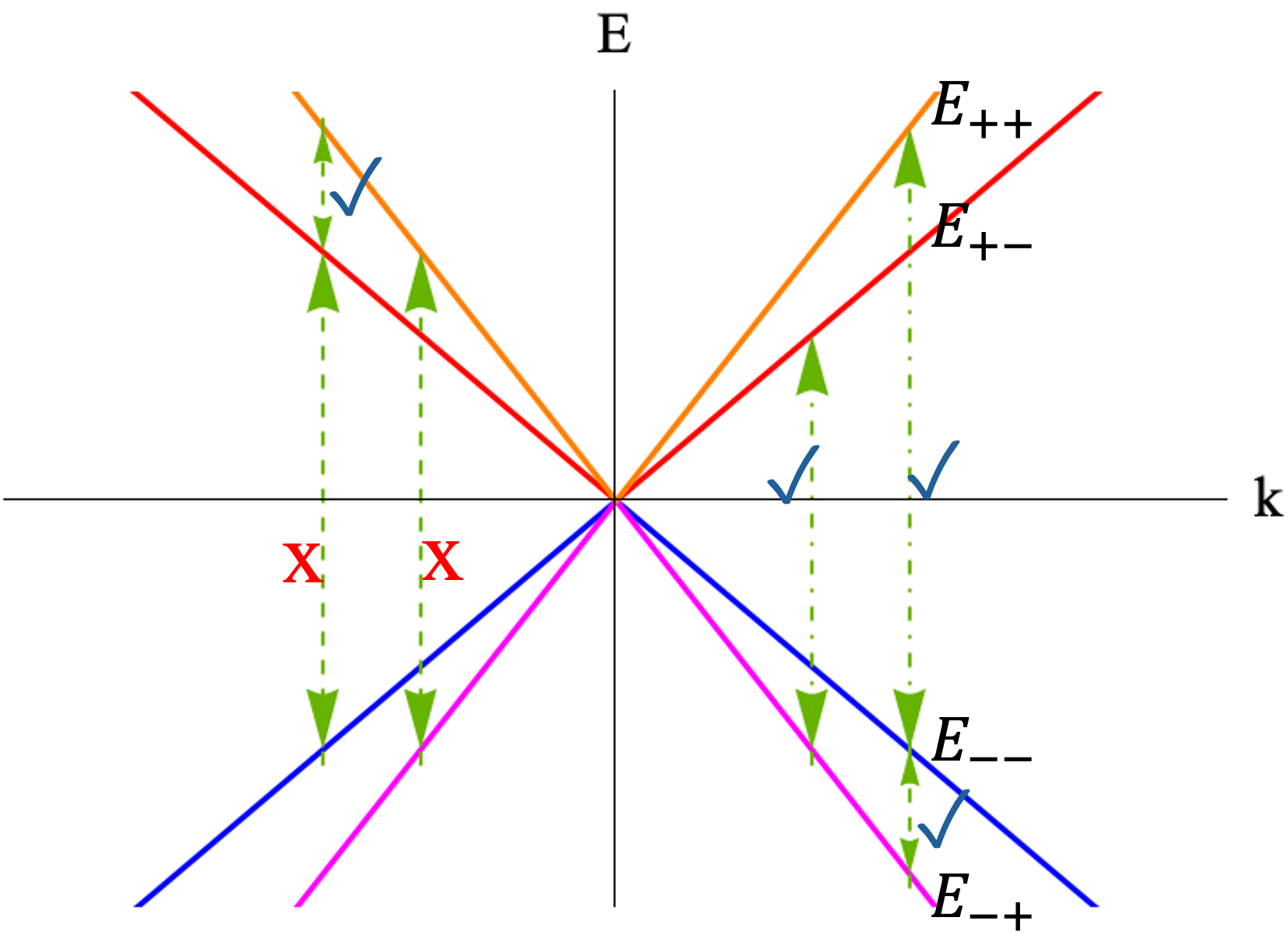}
     \vspace{0.5cm}
  \includegraphics[width=7.5cm,height=5.5cm]{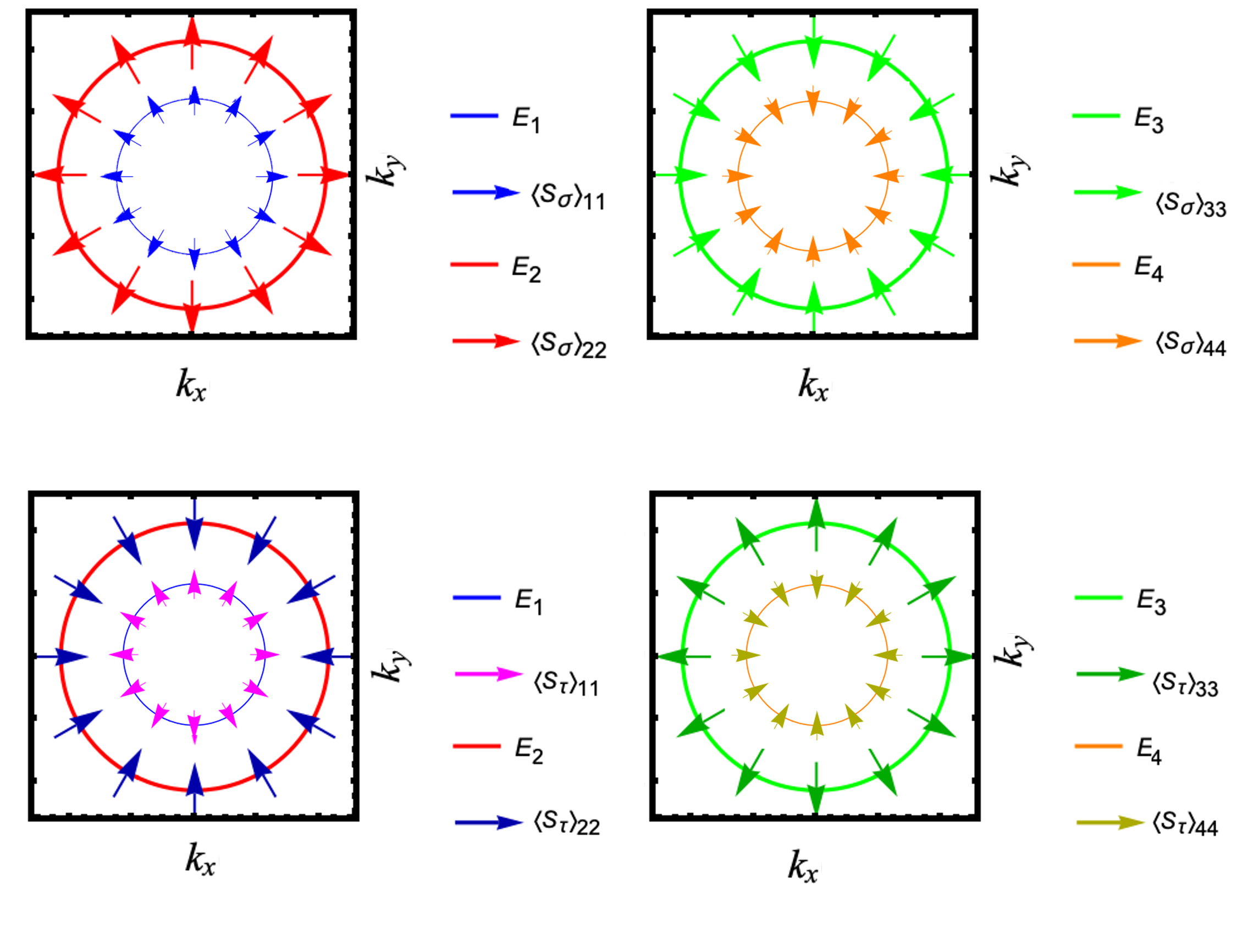}
    \vspace{-0.25cm}
    \caption{(Colour online) Energy dispersion diagram of graphene Kek-Y Hamiltonian (left-hand). The symbols $\surd$ in blue and $\bf \times$ in red indicate allowed and forbidden  \textit{Zitterbewegung} frequencies (spin-transitions), respectively. Note that the electron-hole symmetric bands ($E_{++}\leftrightarrow E_{-+}$ and $E_{+-}\leftrightarrow E_{--}$)  do not present \textit{Zitterbewegung} transition amplitudes between them. (Right-hand panels) pseudospin textures for each band at a fix electon/hole energy.  (Top) Sublattice pseudospin textures. (Bottom) valley pseudospin textures. }
    \label{fig:grapkek}
\end{figure}

There are two main Kekul\'e structural deformations in graphene. They are the Kekul\'e-O and the Kekul\'e-Y bond distortion. In the former, there is an alternate double bond connection between the carbon atoms, just as it occurs in benzene rings.
In the latter however, the Kekul\'e-Y (Kek-Y) bond distortion  involves double bonds in the `Y' shape; this distortion results in the coupling of the two distinct Dirac valleys.
%\cite{cheianov2009hidden}.
The Hamiltonian that describes \textit{low-energy electronic excitations} in graphene with Kek-Y texture bonding was derived by Gamayun et al. \cite{Gamayun}. The Kek-Y Hamiltonian reads,
\begin{equation}\label{eq:kekyham}
    H_{\text{KY}}(\bm k) = v_\sigma\left({\boldsymbol{k}}\cdot\boldsymbol{{S}}_\sigma  \right)+v_\tau\left({\boldsymbol{k}}\cdot \boldsymbol{{S}}_\tau  \right),
\end{equation}

\noindent where  $\boldsymbol{{S}}_\sigma = \tau_0 \otimes \boldsymbol{{\sigma}}$ and $\boldsymbol{{S}}_\tau =   \boldsymbol{{\tau}} \otimes\sigma_0$ (here $N_s=2$),
with coupling constants $\alpha_{1}=v_{\sigma}$ and $\alpha_{2}=v_{\tau}$, describing the Fermi velocities (in units of $\hbar$) respectively, and
$\boldsymbol{{\sigma}} =({\sigma}_x,{\sigma}_y,{\sigma}_z)$ denoting that the pseudospin and $\boldsymbol{{\tau}} =({\tau}_x,{\tau}_y,{\tau}_z)$ is the valley pseudospin. Here, ${\sigma}_{x,y,z}$ and ${\tau}_{x,y,z}$ represent the Pauli matrices in the pseudospin and valley basis, respectively, while $\sigma_0$,$\tau_0 $ are $2\times 2$ identity matrices \cite{A.SantaCruz}. The valley velocity is typically much smaller than the subalattice velocity,  $v_\tau = \Delta_0v_\sigma$, where $v_F = 10$\,\AA$/\rm{s}$~\cite{Gamayun}, is the unperturbed graphene Fermi velocity, and $\Delta_0$ is a dimensionless Kekul\'e coupling strength parameter with the values ranging from 0 to 0.1. Note that the valley locking term is responsible for breaking the valley degeneracy between the $K$ and $K'$ Dirac cones, which eventually produces Dirac cones with different Fermi velocities \cite{A.SantaCruz,Gamayun}. The corresponding band dispersions are given by,

\begin{equation}
 E_{\mu\nu}= \mu(v_{\sigma}+\nu v_{\tau})k, \quad \quad k=\sqrt{k_x^2+k_y^2} , 
\end{equation}

\noindent with $\mu=\pm$ for the electron/hole branch and $\nu=\pm$ associated with the two distinct valleys ($K,K'$), whereas the eigenstates of the Hamiltonian are as follows:
\begin{equation}
    |\psi_{+\nu} \rangle = \frac{1}{2}\begin{pmatrix}
        \nu \re^{-\ri\varphi} \\
    \nu \\
        1 \\
        \re^{\ri\varphi}
    \end{pmatrix}, \quad
    |\psi_{-\nu} \rangle = \frac{1}{2}\begin{pmatrix}
        -\nu \re^{-\ri\varphi} \\
        \nu \\
        -1 \\
        \re^{\ri\varphi}
    \end{pmatrix},
\end{equation}

\noindent with $\varphi = \tan^{-1}(k_y/k_x)$. The energy dispersions  $E_1=E_{++}$, $E_2=E_{+-}$, $E_3=E_{--}$, and $E_4=E_{-+}$ are depicted in figure \ref{fig:grapkek}. 

Now we apply  the formulation outlined in section~\ref{Fifth}. First, note that this four band system gives six possible frequencies $\omega_{ab}=|E_a-E_b|/\hbar$; however, only four of them are independent, namely,  $\omega_{12}=\omega_{34}\equiv\omega_{\tau}=2kv_{\tau}$, $\omega_{13}=\omega_{24}\equiv \omega_{\sigma}=2kv_{\sigma}$, $\omega_{23}=2k(v_{\sigma}-v_{\tau})$, and $\omega_{14}=2k(v_{\sigma}+v_{\tau})$. Then, if we were to analyze the carrier dynamics, one would expect four distinct \textit{Zitterbewegung} modes of oscillations for this system. Nevertheless, this will not be the case here, as two of these frequencies will not be present as we show below. 

In this case $h(\bm k)=0$, whereas 
%due to the electron-hole symmetry between the bands $E_2$ and $E_3$, as well as between bands $E_1$ and $E_4$, it follows that on

\begin{equation}
   \nabla_{\boldsymbol{k}}\left( 0-\frac{1}{2}(E_a + E_b)\right) = 
   \left\{ \begin{array}{lcl}
        v_{\sigma}    & \text{for} & (a=1, b=2)\\
           0          & \text{for} & (a=2, b=3) \quad  \text{and} \quad  (a=1, b=4), \\
         v_{\tau}     & \text{for} &   (a=1, b=3)
   \end{array} \right. 
\end{equation}

\noindent and together with the calculation of the matrix elements of the sublattice pseudospin $\boldsymbol{S}_{\sigma}$ and valley isospin $\boldsymbol{S}_{\tau}$ (see table  \ref{tab:keksum})  leads to the fulfillment of condition (\ref{ZBcondition}) but only for  the transition energies $E_1 \leftrightarrow E_4$ and $E_2 \leftrightarrow E_3$, which are associated with the frequencies $\omega_{14}$ and  $\omega_{23}$, respectively. Therefore, it entails that these frequencies will correspond to vanishing \textit{Zitterbewegung} oscillations. What is remarkable is that such forbidden beating frequencies  are  found without the need of analyzing the full electronic dynamics of the system. Indeed, as recently shown in a study of electronic wave packet dynamics in Kek-Y graphene \cite{A.SantaCruz}, these two frequencies $\omega_{14}$ and  $\omega_{23}$, are absent in the  \textit{Zitterbewegung}  beatings of the time-evolution calculation of the average position and velocity in Kek-Y graphene. The only two beating frequencies allowed in Kek-Y graphene according to our methodology are $\omega_{\sigma}$ and $\omega_{\tau}=\Delta_0 \omega_{\sigma}$, which are in agreement with the numerical approach \cite{A.SantaCruz}. Our results are also in agreement with a recent study of the electronic and optical conductivity in Kek-Y graphene \cite{Naumis}. In this paper the authors find the absence of the transitions between the ``slow'' velocity bands $S_+ \leftrightarrow S_-$ (corresponding to our $E_2 \leftrightarrow E_3$ transitions) and those of the ``fast'' velocity bands  $F_+ \leftrightarrow F_-$ (corresponding to our $E_1 \leftrightarrow E_4$ transitions). The absence of these transitions is understood there in terms of the Fermi golden rule, which gives zero amplitude probabilities for such transitions.  
\begin{table} [!t]
\caption{Table of all matrix elements for the sublattice pseudospin and valley isospin transition amplitudes and pseudospin textures in graphene with Kek-Y bond distortion.}
\vspace{0.2cm}
\centering
\scalebox{0.68}{
\begin{tabular}{|c||c|c|c|c|c|c|}
\hline
 Matrix element / $a\leftrightarrow  b$& $1 \leftrightarrow 2$& $1 \leftrightarrow 3$& $1 \leftrightarrow 4$& $2 \leftrightarrow 3$& $2 \leftrightarrow 4$& $3 \leftrightarrow 4$\\
\hline
\hline
$\langle a| i \mathcal{S}_{\sigma} | b \rangle$ 
 & $(0,0,0)$ 
 & $(\sin \varphi, -\cos \varphi, 0)$ 
 & $(0,0,0)$
 & $(0,0,0)$
 & $(-\sin \varphi, \cos \varphi, 0)$
 & $(0,0,0)$ \\
\hline
$\langle a| i \mathcal{S}_{\tau} | b \rangle$ 
 & $(\sin \varphi, -\cos \varphi, 0)$ 
 & $(0,0,0)$ 
 & $(0,0,0)$
 & $(0,0,0)$
 & $(0,0,0)$
 & $(-\sin \varphi, \cos \varphi, 0)$ \\
\hline
$\langle a| i(\mathcal{S}_{\sigma}+\mathcal{S}_{\tau}) | b \rangle$ 
 & $(\sin \varphi, -\cos \varphi, 0)$ 
 & $(\sin \varphi, -\cos \varphi, 0)$ 
 & $(0,0,0)$
 & $(0,0,0)$
 & $(-\sin \varphi, \cos \varphi, 0)$
 & $(-\sin \varphi, \cos \varphi, 0)$ \\
\hline
$\langle \mathcal{S}_{\sigma} \rangle_{aa} + \langle \mathcal{S}_{\sigma} \rangle_{bb}$& $2(\cos \varphi, \sin \varphi, 0)$ 
 & $(0,0,0)$ 
 & $(0,0,0)$
 & $(0,0,0)$
 & $-2(\cos \varphi, \sin \varphi, 0)$
 & $-2(\cos \varphi, \sin \varphi, 0)$ \\
\hline
$\langle \mathcal{S}_{\tau} \rangle_{aa} + \langle \mathcal{S}_{\tau} \rangle_{bb}$& $(0,0,0)$ 
 & $2(\cos \varphi, \sin \varphi, 0)$ 
 & $(0,0,0)$
 & $(0,0,0)$
 & $(0,0,0)$
 & $(0,0,0)$ \\
\hline
\end{tabular}
}
\label{tab:keksum}
\end{table}

\subsection{Dirac-Weyl Hamiltonians: graphene $\alpha-T_3$ case}

The $\alpha-T_3$ Hamiltonian describes a honeycomb lattice with two sites per unit cell with hopping amplitude $t$ (as graphene) but with extra (carbon) sites at the center of each hexagon and hopping amplitude $\alpha t$. Hence, the  model interpolates between graphene ($\alpha =0$) and the dice lattice ($\alpha=1$) depending upon the value of $\alpha$ \cite{Raoux}. The general low energy effective Hamiltonian for the $\alpha-T_3$ model reads,

\begin{equation}
    H_{1T_3}(\bm k) = v_F \left(\boldsymbol{k}\cdot {\boldsymbol{S}}\right),
\end{equation}

\noindent where $v_F=3ta_o/\sqrt{2}$ is the usual graphene Fermi velocity, $a_o$ is the lattice constant, ${\bm k} = (k_x,k_y,0)$ is the wave vector in the plane, and the pseudospin vector
${\boldsymbol{S}} = \left({S}_x,{S}_y,{S}_z\right)$ (in units of $\hbar$) reads, 

\begin{equation}
\label{Smatrices}
  S_x=\left( \begin{array}{ccc}
      0 & \cos\phi & 0 \\
       \cos\phi & 0 & \sin\phi \\
      0 & \sin\phi & 0
  \end{array}
  \right), \,\,\,\,  
  S_y=\left( \begin{array}{ccc}
      0 & -\ri \cos\phi &  0 \\
      \ri \cos\phi &  0 & -\ri \sin\phi\\
      0 &  \ri \sin\phi &  0
  \end{array}
  \right), \,\,\,\,  S_z=\left( \begin{array}{ccc}
      1 & 0 & 0 \\
      0 & 0 & 0 \\
      0 & 0 & -1
  \end{array}
  \right),
\end{equation}

\noindent with $\phi = \arctan \alpha$, satisfying $\cos \phi = \frac{1}{\sqrt{1 + \alpha^2}}$,   $\sin \phi = \frac{\alpha}{\sqrt{1 + \alpha^2}} $, whereas the low energy dispersions of this three-band model are simply:

\begin{equation}
     E_{0}=0,  \quad E_{\pm}=\pm v_F k ,
\end{equation}

\noindent hence, the main difference between the $\alpha-T_3$ model and pristine graphene bands is the extra flat band sitting at the Fermi level $E_o$. The  eigenvectors are given by, 

\begin{equation}
   |\Psi_0\rangle = 
\begin{pmatrix}
-\re^{-\ri\theta}\sin \phi \\
0 \\
\re^{\ri\theta} \cos \phi
\end{pmatrix}, \quad\quad
     |\Psi_{\pm}\rangle = \frac{1}{\sqrt{2}}
\begin{pmatrix}
\re^{-\ri\theta} \cos \phi \\
\pm1 \\
\re^{\ri\theta} \sin \phi
\end{pmatrix},
\end{equation}
\begin{figure}[!t]
    \centering
     \includegraphics[width=6.0cm,height=6.5cm]{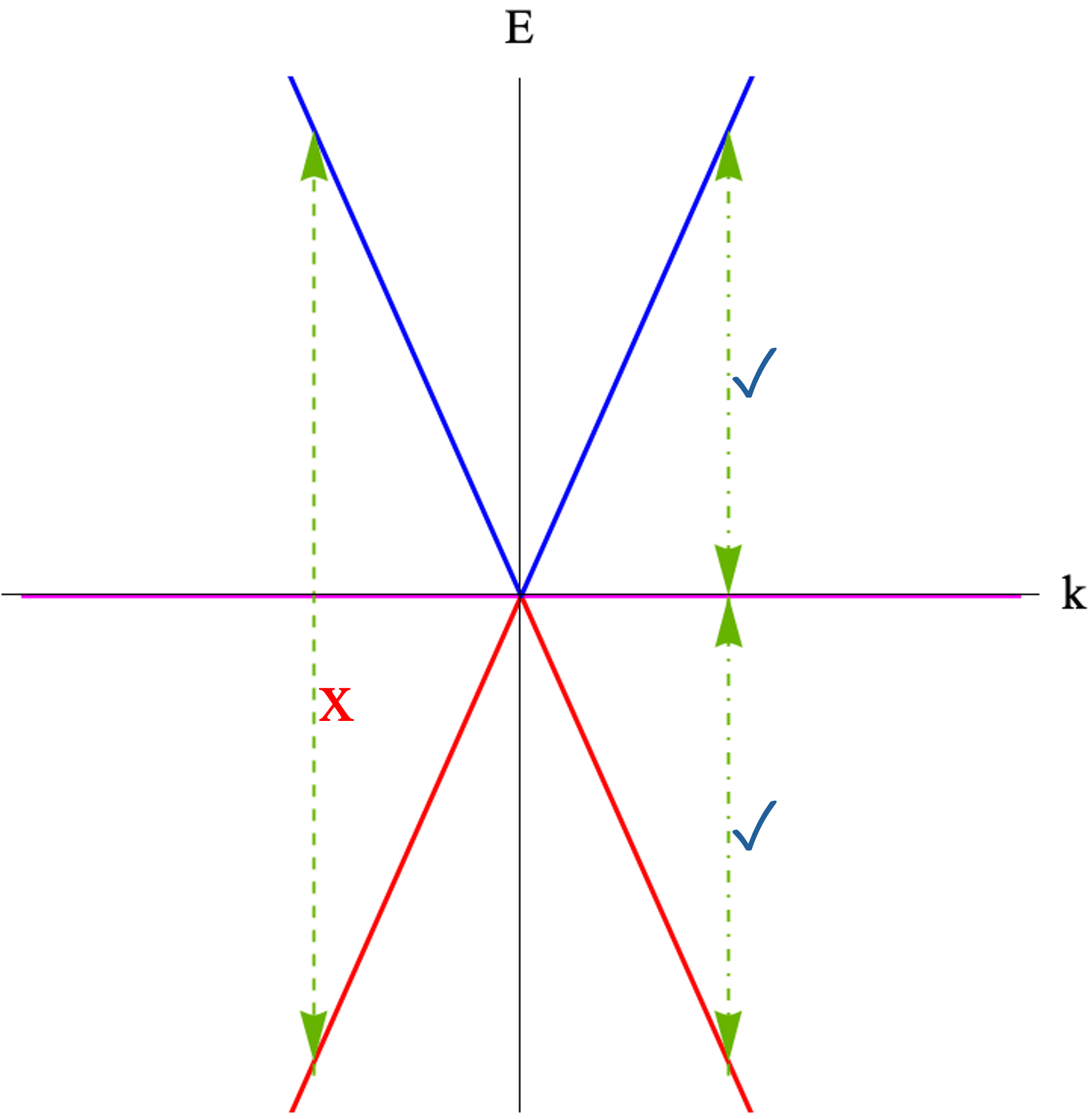}
    \includegraphics[width=0.8\textwidth]{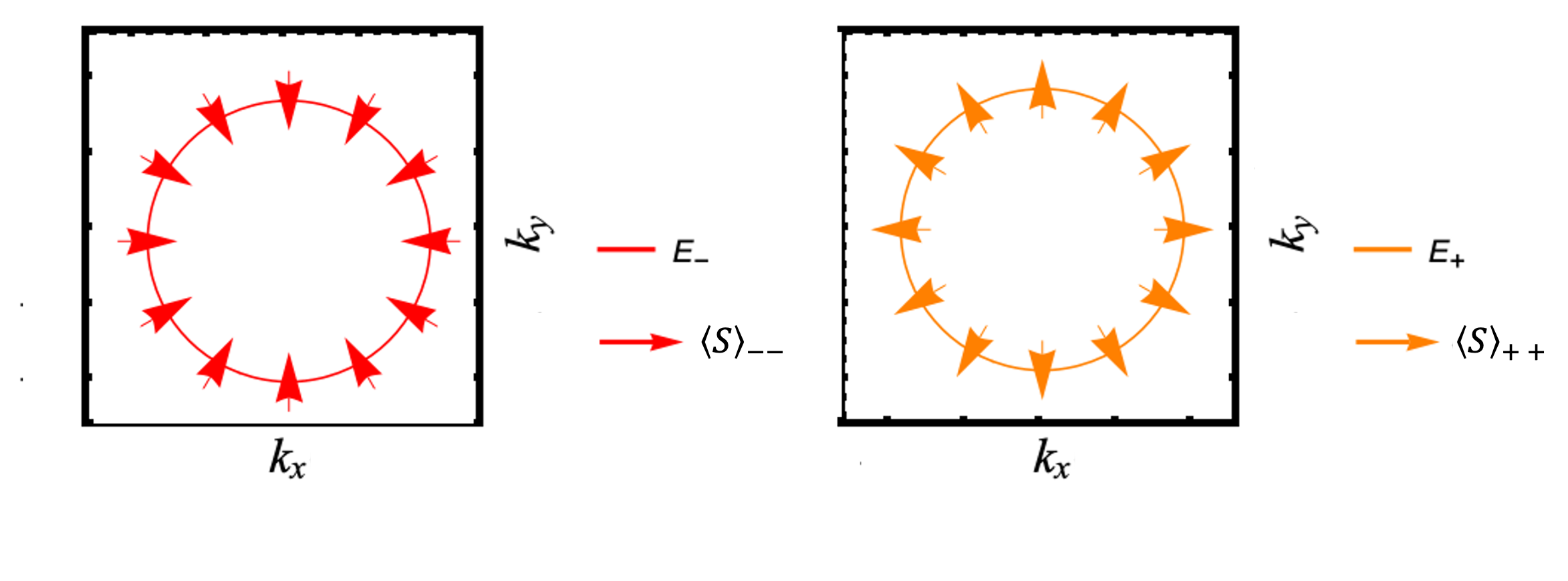}
    \caption{(Colour online) (Top) Band dispersion for the $\alpha-T_3$ model Hamiltonian with $\alpha=1$. Here, the \textit{Zitterbewegung} transition amplitudes that are allowed are only these between the electon/hole band to the zero energy flat band $E_0$. Diagrams illustrate the contour plots of the $E_-$ and $E_+$ bands at a fixed electron/hole energy, respectively. The behavior of  its pseudospin textures are also shown.  
}
    \label{fig:spinalpha}
\end{figure}
\begin{table}[!t]
\caption{Pseudospin transition amplitudes and pseudospin textures of the $\alpha-T_3$ model for $\alpha=1$.}
\vspace{0.2cm}	
\centering
\scalebox{1.0}{
\begin{tabular}{|c||c|c|c|}
\hline
Matrix element / $a\leftrightarrow b$
& $+\leftrightarrow -$ 
& $+\leftrightarrow 0$ 
& $0 \leftrightarrow -$ \\
\hline
\hline
$\langle a| \ri{\bm S}| b \rangle$ 
 & $(0,0,0)$ 
 & $\frac{1}{\sqrt{2}}(-\sin \phi, \cos \phi, -\ri)$ 
 & $\frac{1}{\sqrt{2}}(\sin \phi, -\cos \phi, -\ri)$ \\
\hline
$\langle {\bm S} \rangle_{aa} + \langle {\bm S} \rangle_{bb}$ 
 & $(0,0,0)$ 
 & $(\cos \phi, \sin \phi, 0)$ 
 & $-(\cos \phi, \sin \phi, 0)$ \\
\hline
\end{tabular}
}
\label{table3}
\end{table}

\noindent with $\theta = \arctan(k_y/k_x)$. For the case of $\alpha=1$, the matrices (\ref{Smatrices}) reduce to the spin-1 matrices. Here, two distinct frequencies are expected to be associated to the \textit{Zitterbewegung} oscillations. They are $\omega_{+-}=\omega_{-+}=2v_F k$ and $\omega_{+0}=\omega_{-0}=v_F k$. Then, one naturally would expect that these two frequencies will participate in the \textit{Zitterbewegung} oscillations once the carrier dynamics is performed. 

However, this model yields,  
\begin{equation}
   \nabla_{\boldsymbol{k}}\left( 0-\frac{1}{2}(\mathcal{E}_a + \mathcal{E}_b)\right) = \left \{ \begin{array}{lcl}
        0 & \mbox{for} &   (a=+, b=-) \quad \mbox{and} \quad  (a=-, b=+)\\
     v_{F} & \mbox{for} &   (a=+, b=0) \quad \mbox{and} \quad  (a=-, b=0),  
   \end{array}\right.
   \label{alpha3}
\end{equation}

\noindent and once we calculate the matrix elements of the  pseudospin $\boldsymbol{S}$, (shown in table \ref{table3}), it is evident that the condition (\ref{ZBcondition}) for suppression of the  \textit{Zitterbewegung} amplitudes is satisfied for the frequency $\omega_{+-}$ alone, which is associated with electron-hole transition between $E_+ \leftrightarrow E_-$. This is in contrast to what will occur in pristine graphene.  On the other hand, the condition (\ref{ZBcondition}) does not hold for the frequency $\omega_{+0}=\omega_{-0}$, indicating that this will be the only  \textit{Zitterbewegung} oscillation permitted in the $\alpha-T_3$ model. 

Therefore, for $\alpha=1$ (dice lattice), there is a flat band that controls both its dynamical and optical behavior, as the pseudospin matrix elements indicate that only flat band transitions ($E_{\pm}\leftrightarrow E_0$) contribute to the carrier dynamics, whereas the direct inter-cone transition ($E_+\leftrightarrow E_-$) is forbidden. Hence, the only surviving \textit{Zitterbewegung} frequency is $\omega_{\pm0}=v_Fk$, while the frequency mode $\omega_{+-}=2v_Fk$ is suppressed.
This selection rule matches the optical response reported by Oriekhov and Gusynin \cite{oriekhov}, where the optical conductivity of the dice lattice arises exclusively from flat band to cone transitions, with a single activation threshold at $\omega=v_F|p_z|/\hbar$. Both results show that the flat band is the only active interband channel.

On the other hand, in a recent work by Illes et al. \cite{illes2015} it is shown that the optical conductivity evolves with the Berry phase $\phi_{\text B}=\piup(1-\alpha^2)/(1+\alpha^2)$, from graphene ($\phi_{\text B}=\piup$) to the dice lattice ($\phi_{\text B}=0$). They find that in the graphene case ($\alpha=0$), cone to cone transitions dominate, while flat band to cone transitions dominate for $\alpha=1$, which is consistent with the collapse of the \textit{Zitterbewegung} spectrum to a single $\omega_{\pm0}$ mode. They also report that intermediate $\alpha$ values yield both channels, corresponding to a continuous change in the interband coherence with $\phi_{\text B}$. The suppression of $E_+\leftrightarrow E_-$ transitions found here correspond to the disappearance of the high-energy optical threshold identified by Illes et al. \cite{illes2015}.

Finally, it is important to emphasize that, eventhough for the Dirac-type Hamiltonians studied here, there are both bands with electron-hole symmetry, and there are those bands precisely that exhibit  forbidden \textit{Zitterbewegung} transitions, as such, this is not a necessary condition. Indeed, as shown above, the suppression of the \textit{Zitterbewegung} amplitudes can also arise without such criterion being satisfied, as it to occurs in Rashba-Dresselhaus spin-splitted bands with equal SOC strengths. On the contrary, note also that not all systems that display electron-hole symmetry result in vanishing \textit{Zitterbewegung} frequencies. For example, these are pristine monolayer \cite{Katsnelson}, the bilayer of graphene \cite{RusinZawadski} and graphene with Rashba or with intrinsic SOC. Interestingly,  a link between \textit{Zitterbewegung} and topological phase transitions has been also investigated recently \cite{Shen}.

\section{Summary and conclusions}

In this work, it is shown that spin-related response properties as the  frequency dependent  spin conductivity and the intrinsic spin Hall conductivity of general multiband electronic Hamiltonians are interwoven with the  \textit{Zitterbewegung} amplitudes of such systems. Expressions for the spin related properties revealing such connection were obtained within Kubo linear response formalism.   We also found a useful condition that allows us to directly discern on the presence or not of \textit{Zitterbewegung} oscillations in spin-momentum locked systems and Dirac-type Hamiltonians which only involves a simple analysis of the spin (pseudospin)-transition and spin (pseudospin) textures of these systems. The advantage of this alternative approach is that it permits us to determine whether certain \textit{Zitterbewegung} oscillations frequencies are forbidden in these systems without turning to the actual full quantum mechanical dynamical analysis. We have shown three illustrative examples of the applicability of our approach that manifest the suppression of specific \textit{Zitterbewegung} oscillations. Namely, the joint Rashba-Dresselhaus Hamiltonian for the situation of equal spin-orbit coupling strengths, the case of graphene with Kekul\'e-Y bond distortion, and the dice lattice of the $\alpha-T_3$ model.

\vspace{1cm}
\appendix

\section{Derivation of the static intrinsic Spin Hall conductivity formula}\label{Eq3.6Derivation}

We start by substituting  (\ref{Qij}) into equation~(\ref{SpinC}) for a two-dimensional system with $i\neq j$, 

\begin{equation}
 \sigma^l_{ij}(\omega) =-\frac{e}{2\hbar^3\omega}\frac{1}{A}\sum_{\mathbf{k}}\sum_{a,b\neq a} \lim_{\delta \rightarrow 0^+} K_{ba} (\hbar \omega+\ri\delta)   (E_a - E_b) \text{Tr}  \left [ \frac{\hbar}{2} \left \{ \sigma_l , v_i\right\}   \hat Z_j^{ab}  \right ],
 \end{equation}

\noindent now, since $\hat Z_j^{ab} = \ri(E_b -E_a)^{-1}Q_a \hbar v_j Q_b$ for $a\neq b$ and ${\cal J}_i^l =\frac{\hbar}{4}\{\sigma_l,v_i\}$, after adding a small positive imaginary part $\eta$ to the frequencies $\omega$ \cite{Marder}, { i.e.,} $\hbar\omega \rightarrow \hbar\omega + \ri\eta= \hbar \bar \omega$, and writing explicitly the trace, we get

\begin{equation}
 \sigma^l_{ij}(\omega) =\frac{\ri e}{\hbar^2\bar \omega}\frac{1}{A} \sum_{\mathbf{k}}\sum_{a}\sum_{b\neq a}  \frac{n_F(E_a) - n_F(E_b)}{E_a - E_b+\hbar\bar \omega}\langle b|{\cal J}_i^l|a\rangle\langle a|v_j|b\rangle,
\end{equation}
which can be expressed as,
\begin{equation}
 \sigma^l_{ij}(\omega) =\frac{\ri e}{\hbar}\frac{1}{A} \sum_{\mathbf{k}} \left ( \sum_{a}\sum_{b\neq a}  \frac{n_F(E_a)\langle b|{\cal J}_i^l|a\rangle\langle a|v_j|b\rangle}{\hbar\bar \omega(E_a - E_b+\hbar\bar \omega)}  -\sum_{a}\sum_{b\neq a} \frac{n_F(E_b)\langle b|{\cal J}_i^l|a\rangle\langle a|v_j|b\rangle}{\hbar\bar \omega(E_a - E_b+\hbar\bar \omega)} \right) ,
\end{equation}

\noindent and by interchanging $a\leftrightarrow b$ in the second term and using the expansion

\begin{equation}
    \frac{1}{E_a - E_b\pm\hbar\bar \omega}=\frac{1}{E_a - E_b}\left(1\mp\frac{\hbar\bar \omega}{E_a - E_b} \right)+{\cal O}(\bar\omega^2),
\end{equation}

\noindent the spin-Hall optical conductivity can be written as $\sigma^l_{ij}(\omega) = \sigma^{l_1}_{ij}(\omega)+\sigma^{l_2}_{ij}(\omega)$, with

\begin{equation}
 \sigma^{l_1}_{ij}(\omega) =\frac{\ri e}{\hbar}\frac{1}{A} \sum_{\mathbf{k}}  \sum_{a\neq b}\sum_{b}  \frac{n_F(E_a)}{\hbar\bar \omega(E_a - E_b)}\left ( \langle b|{\cal J}_i^l|a\rangle\langle a|v_j|b\rangle +\langle b|v_j|a\rangle \langle a|{\cal J}_i^l|b\rangle \right) +{\cal O}(\bar\omega^2).
\end{equation}

\noindent Using the Heisenberg's equation of motion $v_i=\frac{1}{\ri\hbar}[x_i,H]$,  it follows that $\langle b|v_j|a\rangle=-\langle a|v_j|b\rangle=\frac{1}{\ri\hbar}(E_a-E_b)\langle b|x_j|a\rangle$. This result, together with the property $\sum_b Q_b=\sum_b|b\rangle\langle b|=I_N$ and the fact that the commutator $[x_i,{\cal J}^l_j]=0$, since  $[x_i,v_j]=0$ and $[x_i,\sigma_l]=0$ for $i\neq j$, give that the sum of the product of the matrix elements in parenthesis adds up to zero. Therefore, $\sigma^{l_1}_{ij}(\omega)$ is vanishingly small, while in the limit of zero frequency, $\sigma^{l_1}_{ij}(0)=0$.   

\vspace{0.5cm}
  On the other hand, we have
\begin{equation}\label{Sigma2}
 \sigma^{l_2}_{ij}(\omega) =\frac{\ri e}{\hbar}\frac{1}{A} \sum_{\mathbf{k}}  \sum_{a\neq b}\sum_{b}  \frac{n_F(E_a)}{(E_a - E_b)^2}\left ( \langle b|{\cal J}_i^l|a\rangle\langle a|v_j|b\rangle -\langle b|v_j|a\rangle \langle a|{\cal J}_i^l|b\rangle \right) +{\cal O}(\bar\omega^2).
\end{equation}

\noindent which, by noticing that $\langle b|{\cal J}_i^l|a\rangle = \langle a|{\cal J}_i^l|b\rangle^*$ and $\langle a|v_j|b\rangle=\langle b|v_j|a\rangle^*$ since both $v_j$ and ${\cal J}_i^l$ are Hermitian, gives directly that in the limit of zero frequency, 

\begin{equation}
 \sigma^{l_2}_{ij}(0) =\frac{e}{\hbar}\frac{1}{A} \sum_{\mathbf{k}}  \sum_{a}n_F(E_a) \Omega_a^l,
\end{equation}

\noindent where $\Omega_a^l$ is the Berry curvature and is given by 
\begin{equation}
    \Omega_a^l=\sum_{b\neq a} \frac{2\,\text{Im} \langle a|{\cal J}_i^l|b\rangle\langle b|v_j|a\rangle}{(E_a - E_b)^2},
\end{equation}

\noindent hence, the dc intrinsic spin Hall conductivity reads 

\begin{equation}\label{SHC0}
    \sigma^{l}_{ij}(0) = 0+\sigma^{l_2}_{ij}(0). 
\end{equation}

Now, since we are considering periodic systems,  note that the band indices $a$ and $b$ represent  Brillouin zone wave vector pairs $(a,\bm k)$ and $(b,\bm k)$  for the Bloch states. Then, we can adopt a new notation as $|a\rangle\equiv |\psi_a({\bm k})\rangle \rightarrow |\bm k n\rangle$ and $|b\rangle \equiv |\psi_b({\bm k'})\rangle\rightarrow |\bm k'n' \rangle$ with $n,n'$ being the new band indices. Also note that the matrix elements between the $\bm k$ and $\bm k'$ states are assumed to be zero if $\bm k \neq \bm k'$, that is,   $\langle a|{\cal J}_i^l|b\rangle=\langle \bm k n|{\cal J}_i^l|\bm k' n'\rangle\delta_{\bm k\bm k'}$ and $\langle b|v_j|a\rangle=\langle \bm k' n'|v_j|\bm k n\rangle\delta_{\bm k\bm k'}$. In the same notation, the Fermi distribution is $n_a(E_F)\rightarrow f_{\bm k n}$, and the band energies $E_a\rightarrow E_{\bm k n}$. Therefore, by choosing $l=z$, with $i=x$ and $j=y$, the formula for the dc intrinsic spin Hall conductivity is recasted as 

\begin{equation}
    \sigma^{z}_{xy}(0) =\frac{e}{\hbar}\frac{1}{A} \sum_{\bm k}  \sum_{n} f_{\bm k n}  \sum_{n'\neq n} \frac{2\,\text{Im} \langle \bm k n|{\cal J}_x^z|\bm k n'\rangle\langle \bm k n'|v_y|\bm k n\rangle}{(E_{\bm k n} - E_{\bm k n'})^2},
\end{equation}

\noindent which is the standard Kubo formula for the static spin Hall conductivity, see for instance \cite{Guo,Feng}.
%\vspace{0.3cm}

Finally, it is  straightforward to arrive at the formula (\ref{Sigma0}) for the spin conductivity in terms of the \textit{Zitterbewegung} amplitudes $\hat Z_i^{ab}$; the procedure is as follows. Starting from equations~(\ref{Sigma2}) and~(\ref{SHC0}), interchange the summation indices ($b\leftrightarrow a$) in the second term  of (\ref{Sigma2}) and write the full expression of the spin conductivity in terms of the trace on the band indices.  Then, substitute back the explicit expressions for $v_j=\frac{1}{\hbar} \frac{\partial H}{\partial k_i}$, for the spin current operator ${\cal J}_i^l$, and use once again the identity, $
\hat Q_a \frac{\partial H}{\partial k_i} \hat Q_b = \delta_{ab} \frac{\partial E_a}{\partial k_i} \hat Q_a + (E_b - E_a) \hat Z_i^{ab}$ for $a\neq b$. This leads directly to formula (\ref{Sigma0}).

\section{Connection formula of the \textit{Zitterbewegung} with the spin-textures}
\label{Eq5.2Derivation}

In this appendix we outline the derivation of the formula~(\ref{eq:proof}). We start with an Hermitian Hamiltonian $H(\bm k)$ that describes a nondegenerate multiband system, such as, $ H(\bm k)|\psi_a({\bm k})\rangle =E_a|\psi_a({\bm k})\rangle$, where $E_a$ is the eigenvalue corresponding to the eigenket $|\psi_a({\bm k})\rangle$, with $\langle\psi_a({\bm k})  |\psi_b({\bm k})\rangle = \delta_{ab}$. In what follows, to simplify the notation we use $|\psi_a({\bm k})\rangle \rightarrow |a\rangle$ and drop the $\bm k$-dependence everywhere. Let us now introduce the linear combination $|\lambda_\pm \rangle = |a\rangle\pm \ri|b\rangle$, with $a\neq b$. Then, $ H|\lambda_+\rangle= E_a|a\rangle+\ri E_b|b\rangle$, and therefore, 
\begin{equation}\label{Ap1}
     \nabla_{\boldsymbol{k}}H|\lambda_{+}\rangle  =\left(\nabla_{\boldsymbol{k}}E_a\right)|a\rangle+ E_a\nabla_{\boldsymbol{k}}|a\rangle  
+\ri\left(\nabla_{\boldsymbol{k}}E_b\right)|b\rangle +\ri E_b\nabla_{\boldsymbol{k}}|b\rangle .
\end{equation}

In the same notations, note that

\begin{equation}\label{Ap2}
     \nabla_{\boldsymbol{k}}H|\lambda_{+}\rangle  =\left(\nabla_{\boldsymbol{k}}H\right) (|a\rangle+\ri|b\rangle) +  H\nabla_{\boldsymbol{k}}(|a\rangle+\ri|b\rangle).
\end{equation}

Next, consider a generic multiband Hamiltonian with a pseudospin momentum lock term of the form~(\ref{eq:con}), 
\begin{equation} 
    H(\bm k) =h(\boldsymbol{\bm k})+\sum_{\eta=1}^{N_s} \alpha_\eta \left(\bm k \cdot \boldsymbol{\mathcal S}_{\eta}\right),
    \nonumber
    %,\hspace{0.5cm} \eta =\{s,\sigma,\tau}\}
\end{equation}

\noindent which leads directly to
\begin{equation}\label{Ap3}
    \nabla_{\boldsymbol{k}}\hat{H} =     \nabla_{\boldsymbol{k}}h(\boldsymbol{k})+ \sum_{\eta =1}^{N_s} \alpha_\eta {\boldsymbol{\mathcal{S}}}_{\eta}. 
\end{equation}

We then substitute \ref{Ap3} into the right-hand side of \ref{Ap2}  and calculate two expressions for $ \langle \lambda_{\pm}|\nabla_{\boldsymbol{k}}H|\lambda_{+}\rangle $ using both  \ref{Ap1} and \ref{Ap2}.  After simple algebra we arrive at the identity

\begin{equation}
\begin{aligned}
     \sum_{\eta=1}^{N_s}\left[\alpha_\eta\left(\langle{\boldsymbol{\mathcal{S}}}_{\eta}\rangle_{aa} \pm \langle{\boldsymbol{\mathcal{S}}}_{\eta}\rangle_{bb}+\ri\langle a|{\boldsymbol{\mathcal{S}}}_{\eta}|b\rangle \mp \ri\langle b|{\boldsymbol{\mathcal{S}}}_{\eta}|a\rangle\right)\right]
         +\nabla_{\boldsymbol{k}}h(\boldsymbol{k})& \\= 
        \nabla_{\boldsymbol{k}} \left(E_a\pm E_b\right)+ \left(E_b-E_a\right)\left(\langle a|\ri\nabla_{\boldsymbol{k}}|b\rangle \pm \langle b|\ri\nabla_{\boldsymbol{k}}|a\rangle\right),
        \end{aligned}
\end{equation}

\noindent where $\langle{\boldsymbol{\mathcal{S}}}_{\eta}\rangle_{a a}\equiv\langle a|\boldsymbol{\mathcal S}_{\eta}|a\rangle$. Hence, for the $\langle \lambda_+|$ case we have,

\begin{equation}\label{eq:proof5}
    \begin{aligned}
&\sum_{\eta=1}^{N_s}\alpha_\eta\Bigl(\langle{\boldsymbol{\mathcal{S}}}_{\eta}\rangle_{aa}+ \langle{\boldsymbol{\mathcal{S}}}_{\eta}\rangle_{bb}+2\Re{\left[\ri\langle a|{\boldsymbol{\mathcal{S}}}_{\eta}|b\rangle\right]}\Bigr)\\
         &+\nabla_{\boldsymbol{k}}h(\boldsymbol{k})= 
        \nabla_{\boldsymbol{k}} \left(E_a+ E_b\right)+2\left(E_b-E_a\right)\Re\left(\langle a|\ri\nabla_{\boldsymbol{k}}|b\rangle \right),
    \end{aligned}
\end{equation}
where $\Re[\ldots]$ denotes the real part. Note that the first two terms  are nothing else but the usual \textit{spin-orientation} definition, while the last term to the right is the real part of the \textit{Berry connection matrix}. Equation \ref{eq:proof5} could be rewritten as,  
\begin{align}
    \Re\left(\langle a|\ri\nabla_{\boldsymbol{k}}|b\rangle \right) &=  \frac{1}{2(E_b-E_a)}\Biggl(\sum_{\eta=1}^{N_s}\alpha_\eta\left(\langle{\boldsymbol{\mathcal{S}}}_{\eta}\rangle_{aa}+ \langle{\boldsymbol{\mathcal{S}}}_{\eta}\rangle_{bb}+2\Re{\left[\ri\langle a|{\boldsymbol{\mathcal{S}}}_{\eta}|b\rangle\right]}\right)\nonumber \\
   & -\nabla_{\boldsymbol{k}} \left(E_a+ E_b-2h(\bm k)\right)\Biggr)
    \label{Ap6}
\end{align}

\noindent while for the $\langle \lambda_-|$ case, we have,

\begin{align}
   % \begin{aligned}
\sum_{\eta=1}^{N_s}\alpha_\eta\Bigl(\langle{\boldsymbol{\mathcal{S}}}_{\eta}\rangle_{aa}&- \langle{\boldsymbol{\mathcal{S}}}_{\eta}\rangle_{bb}+2\text{Im}{\left[\ri\langle a|{\boldsymbol{\mathcal{S}}}_{\eta}|b\rangle\right]}\Bigr)
         +\nabla_{\boldsymbol{k}}h(\boldsymbol{k})\nonumber\\
        &= 
        \nabla_{\boldsymbol{k}} \left(E_a- E_b\right)+2\ri\left(E_b-E_a\right)\text{Im}\left(\langle a|\ri\nabla_{\boldsymbol{k}}|b\rangle \right)
   % \end{aligned}
\label{eq:proof5}
\end{align}
where $\text{Im}[\cdot\cdot\cdot]$ denotes the imaginary part. Given that $\langle{\boldsymbol{\mathcal{S}}}_{\eta}\rangle_{a,b}$, $E_a$ and $E_b$ are real parameters, then we identify 
\begin{equation}\label{eq:proof5}
    \text{Im}\left(\langle a|\ri\nabla_{\boldsymbol{k}}|b\rangle \right) =  \frac{1}{(E_b-E_a)}\sum_{\eta=1}^{N_s}\alpha_\eta\text{Im}{\left[\ri\langle a|{\boldsymbol{\mathcal{S}}}_{\eta}|b\rangle\right]}, 
\end{equation}

\noindent and since in general we can write
\begin{equation}\label{eq:proof10}
    \langle a|\ri\nabla_{\boldsymbol{k}}|b\rangle   = \Re{\left[\langle a|\ri\nabla_{\boldsymbol{k}}|b\rangle\right]}+\ri\,\text{Im}\left[\langle a|\ri\nabla_{\boldsymbol{k}}|b\rangle \right]  ,
\end{equation}
finally, using  \ref{Ap6} for the real part and  \ref{eq:proof5} for the imaginary part, it produces the interconnection between \textit{Zitterbewegung} and spin texture and spin-transition amplitudes shown in equation~(\ref{eq:proof}).

\section*{Acknowledgements}
  We acknowledge the support of DGAPA-UNAM through the project PAPIIT No. IN111624.

\ukrainianpart

\title{Зв'язок явища {\textit{Zitterbewegung}} зі спіновою провідністю та спіновими текстурами багатозонних систем}

\author
{Ф. Мірелес\refaddr{label1},
	Е. Ортіз
		\refaddr{label1,label2}}

\addresses{
	\addr{label1} Фізичний факультет, Центр нанотехнологій Національного автономного університету Мехіко (UNAM), п/с~14, 22800 Енсенада, Баха Каліфорнія, Мексика
	\addr{label2} Факультет електротехніки, Університет Нотр-Дам, Нотр-Дам, IN 46556, США
}

\makeukrtitle

\begin{abstract}
	\tolerance=3000%
	Відомо, що явище \textit{Zitterbewegung} у багатозонних електронних системах тонко пов'язане з зарядовою провідністю, кривизною Беррі та числом Черна. Ми показуємо, що деякі спінові характеристики, такі як оптична спінова провідність та власна спінова провідність Холла, також пов'язані з амплітудами \textit{Zitterbewegung}. Також продемонстровано, що в багатозонних гамільтоніанах Дірака можна встановити прямий зв'язок між \textit{Zitterbewegung} зі спіновими текстурами та амплітудами спінових переходів. Останні дозволяють нам розпізнати наявність чи відсутність  \textit{Zitterbewegung} просто аналізуючи спінові або псевдоспінові текстури. Ми наводимо приклади застосовності нашого підходу для гамільтонових моделей, які демонструють придушення специфічних \textit{Zitterbewegung}.

	\keywords спін-орбітальна взаємодія, 2DEGs, спіновий ефект Холла, спінова провідність, явища спінового переносу
\end{abstract}


\begin{thebibliography}{66}

\bibitem{schrodinger1930kraftefreie}
Schr\"odinger E., Sitzungsber. Preuss. Akad. Wiss., Phys. Math KI., 1930, {\bf 24}, 418. 

\bibitem{GeneralTheory-PRB2010}
D\'avid G.,  Cserti J., Phys. Rev. B, 2010, {\bf 81}, 121417, \doi{10.1103/PhysRevB.81.121417}.

\bibitem{CertiDavid2010}
Cserti J.,  D\'avid G., Phys. Rev. B, 2010, {\bf 82}, 201405(R), \doi{10.1103/PhysRevB.82.201405}.

\bibitem{Lurie&Cremer}
Luri\'e D.,  Cremer S., Physica, 1970, {\bf 50}, 224, \doi{10.1016/0031-8914(70)90004-2}.

\bibitem{Cannata}
Cannata F., Ferrari L.,  Russo G., Solid State Commun., 1990, \textbf{74}, 309, \doi{10.1016/0038-1098(90)90192-E}.

\bibitem{SpintronicsReview}
Zuti\'c J., Fabian J.,   Das Sarma S.,  Rev. Mod. Phys., 2004, \textbf{76}, 323, \doi{10.1103/RevModPhys.76.323}.  

\bibitem{Hirohata} 
Hirohata A.,  Yamada K.,  Nakatani Y.,  Prejbeanu  I.-L.,  Di\'{e}ny B.,  Pirro Ph.,  Hillebrands B., J. Magn. Magn. Mater., 2020, \textbf{509}, 166711, \doi{10.1016/j.jmmm.2020.166711}.  

\bibitem{schliemann2005zitterbewegung}
Schliemann J., Loss D.,  Westervelt R. M., Phys. Rev. Lett., 2005, \textbf{94}, 206801, \doi{10.1103/PhysRevLett.94.206801}. 

\bibitem{jiang2005semiclassical}
Jiang Z. F.,   Li R. D.,  Zhang S.-C.,  Liu W. M., Phys. Rev. B, 2005, \textbf{72}, 045201,  \doi{10.1103/PhysRevB.72.045201}.

\bibitem{Zawadski-PRB2005}
Zawadzki W., Phys. Rev. B, 2005, \textbf{72}, 085217, \doi{10.1103/PhysRevB.72.085217}

\bibitem{gerritsma2010quantum} 
Gerritsma R.,  Kirchmair  G.,  Z\"{a}hringer F.,  Solano E.,  Blatt R.,  Roos C. F., Nature, 2010, \textbf{463}, 68, \doi{10.1038/nature08688}.

\bibitem{leblanc2013direct} 
 LeBlanc L. J.,  Beeler M. C.,  Jim\'{e}nez-Garc\'{i}a K.,  Perry A. R.,  Sugawa S.,  Williams R. A.,   Spielman I. B., New J. Phys., 2013, {\bf 15}, 073011, \doi{10.1088/1367-2630/15/7/073011}.

\bibitem{SOCBEC} 
Qu C.,   Hamner C.,  Gong M.,  Zhang C.,  Engels P., Phys. Rev. A, 2013, {\bf 88}, 021604, \doi{10.1103/PhysRevA.88.021604}.

\bibitem{OpticalSimulationZitter}
Silva T. L.,   Taillebois E. R. F.,  Gomes R. M.,  Walborn S. P.,  Avelar A. T., Phys. Rev. A, 2019, {\bf 99}, 022332, \doi{10.1103/PhysRevA.99.022332}.

\bibitem{perovskite} 
Wen W.,   Liang J.,  Xu H.,  Jin F., Rubo Yu. G.,  Liew T. C. H.,  Su R., Phys. Rev. Lett. 2024, {\bf 133}, 116903, \doi{10.1103/PhysRevLett.133.116903}.

\bibitem{Maksimova} 
Maksimova G. M., Demikhovskii V. Y., Frolova E.V., Phys. Rev. B, 2008, {\bf 78}, 235321, \doi{10.1103/PhysRevB.78.235321}.

\bibitem{RusinZawadski} 
Rusin T. M.,  Zawadzki W., Phys. Rev. B, 2007, {\bf 76}, 195439,
\doi{10.1103/PhysRevB.76.195439}.

\bibitem{RusinZawadski2008}
Rusin T. M.,  Zawadzki W., Phys. Rev. B, 2008, {\bf 78}, 125419,
\doi{10.1103/PhysRevB.78.125419}.

\bibitem{Carrillo2018} 
Carrillo Bastos R., Ochoa M, Zavala S. A., 
Mireles F., Phys. Rev. B, 2018, {\bf 98}, 165436, \doi{10.1103/PhysRevB.98.165436}.

\bibitem{A.SantaCruz}
Santacruz A., Iglesias P. E., Carrillo Bastos R.,  Mireles F., Phys. Rev. B, 2022, {\bf 105}, 205405, \doi{10.1103/PhysRevB.105.205405}.

\bibitem{wpd-phosphorene}
Cunha S. M.,  da Costa D. R.,  de Sousa G. O.,  Chaves A.,  Milton Pereira, Jr. J.,   Farias G. A., Phys. Rev. B, 2019, {\bf 99}, 235424, \doi{10.1103/PhysRevB.99.235424}.

\bibitem{wpd-silicene} 
Romera E., Rold\'an J.,  de los Santos F., Phys. Lett. A, 2014, {\bf 378}, 2582, \doi{10.1016/j.physleta.2014.06.040}.

\bibitem{wpd-silicene2} 
Szafran B., Rzeszotarski B.,  Mre\'nca-Kolasi\'nska A., Phys. Rev. B, 2019, {\bf 100}, 085306, \doi{10.1103/PhysRevB.100.085306}.

\bibitem{Hassan}
Hassan A. M.,  Rashid S.,  Manzoor K.,  Riaz N.,  Ali H.,  Ullah A.,  Imtiaz Khan M., Phys. Lett. A, 2025, {\bf 552}, 130655, \doi{10.1016/j.physleta.2025.130655}.

\bibitem{borophene2} 
Yar A.,  Ilyas A., J. Phys. Soc. Jpn., 2020, {\bf 89}, 124705, \doi{10.7566/JPSJ.89.124705}.

\bibitem{Biswas} 
Biswas T.,  Ghosh T. K., J. Phys.: Condens. Matter, 2018, {\bf 30}, 075301, \doi{10.48550/arXiv.1710.04790}.

\bibitem{wpd-ti} 
Demikhovskii V. Y.,  Telezhnikov A., JETP Lett., 2014, {\bf 99}, 104, \doi{10.1134/S0021364014020064}.

\bibitem{wpd-ti1} 
Ferreira G. F.,  Maciel R. P.,  Penteado P. H.,   Egues J. C., Phys. Rev. B, 2018, {\bf 98}, 165120, \doi{10.1103/PhysRevB.98.165120}.

\bibitem{wpd-ti2} 
Yar A.,  Naeem M.,   Khan S. U.,  Sabeeh K., J. Phys.: Condens. Matter, 2017, {\bf 29}, 465002, \doi{10.1088/1361-648X/aa801a}.

\bibitem{PeetersgroupPRL2021} 
Lavor L. R., da Costa D. R.,  Covaci L.,  Milo\v{s}evi\'{c} M. V.,  Peeters F. M.,   Chaves A., Phys. Rev. Lett., 2021, {\bf 127}, 106801, \doi{10.1103/PhysRevLett.127.106801}.

\bibitem{Berry}
Berry M. V., Proc. R. Soc. London, Ser. A,  1984, {\bf 392}, 45, \doi{10.1098/rspa.1984.0023}.

\bibitem{Rashba} 
Rashba E. I., Sov. Phys.-Solid State, 1960, {\bf 2}, 1109. 

\bibitem{Bychkov} 
Bychkov Y. A.,  Rashba E. I., J. Phys. C: Solid State Phys., 1984, {\bf 17}, 6039, \doi{10.1088/0022-3719/17/33/015}.

\bibitem{Santana} 
Santana Suar\'ez E., Mireles F., Condens. Matter Phys., 2023, {\bf 26}, No.~1, 13504, \doi{10.5488/CMP.26.13504}.

\bibitem{Bercioux}
Bercioux D.,  Lucignano P., Rep. Prog. Phys., 2015,  {\bf 78}, 106001, \doi{10.1088/0034-4885/78/10/106001}.

\bibitem{Dresselhaus} 
Dresselhaus G., Phys. Rev., 1955, {\bf 100}, 580, \doi{10.1103/PhysRev.100.580}. 

\bibitem{Luttinger} 
Luttinger J. M., Phys. Rev., 1956, {\bf 102}, 1030, \doi{10.1103/PhysRev.102.1030}.

\bibitem{Wong} Wong A.,  Mireles F., Phys. Rev. B, 2010, {\bf 81}, 085304, \doi{10.1103/PhysRevB.81.085304}. 


\bibitem{Bernevig}
Bernevig B. A., Phys. Rev. B, 2005, {\bf 71}, 073201, \doi{10.1103/PhysRevB.71.073201}.

%\bibitem{Footnote} In Ref. \cite{Bernevig} the energies are measured in units $\hbar^2$, consequently its equation~(6) for the spin conductivity should be divided by $\hbar^2$ to get the right units of $e/\hbar$ for the spin conductivity.  

\bibitem{Gamayun}
Gamayun O. V., Ostroukh V. P.,  Gnezdilov N. V., Adagideli I.,  Beenakker C. W. J., New J. Phys., 2018, {\bf 20}, 023016, \doi{10.1088/1367-2630/aaa7e5}.


\bibitem{Naumis} Herrera S. A., Naumis G. G., Phys. Rev. B, 2020, {\bf 101}, 205413, \doi{10.1103/PhysRevB.101.205413}.

 
\bibitem{Raoux} 
Raoux A., Morigi M., Fuchs J.-N., Piechon F.,  Montambaux G., Phys. Rev. Lett., 2014 {\bf 112}, 026402, \doi{10.1103/PhysRevLett.112.026402}.
 

\bibitem{Sutherland}  
Sutherland B., Phys. Rev. B, 1986, {\bf 34}, 5208, \doi{10.1103/PhysRevB.34.5208}.

\bibitem{Bercioux2} 
Bercioux D., Urban D. F., Grabert H.,  H\" ausler W., Phys. Rev. A, 2009, {\bf 80}, 063603, \doi{10.1103/PhysRevA.80.063603}.

\bibitem{Katsnelson}  
Katsnelson M. I., Eur. Phys. J. B, 2006, {\bf 51}, 157, \doi{10.1140/epjb/e2006-00203-1}.

\bibitem{Shen} 
Shen. X., Zhu Y. Q., Li Z., Phys. Rev. B, 2022, {\bf 106}, L180301, \doi{10.1103/PhysRevB.106.L180301}.

\bibitem{Marder} Marder M., Condensed Matter Physics, John Wiley and Sons, Inc., New York, 2000.

\bibitem{Guo} Guo G. Y., Murakami S., Chen T.-W.,  Nagaosa N., Phys. Rev. Lett. 2008, {\bf 100}, 096401, \doi{10.1103/PhysRevLett.100.096401}.

\bibitem{Feng} Feng W., Liu C.-C., Liu G.-B., Zhou J.-J., Yao Y., Comput. Mater. Sci., 2016, \textbf{112}, 428--447, \doi{10.1016/j.commatsci.2015.09.020}.

\bibitem{oriekhov}
 Oriekhov D. O., Gusynin V. P., Phys. Rev. B, 2022, {\bf 106}, 115143, \doi{10.1103/PhysRevB.106.115143}.

\bibitem{illes2015}
Illes E., Carbotte J. P.,  Nicol E. J., Phys. Rev. B, 2015, {\bf 92}, 245410, \doi{10.1103/PhysRevB.92.245410}.


\end{thebibliography}
\end{document}